\RequirePackage{fix-cm}
\documentclass[smallextended]{svjour3}       
\smartqed 
\usepackage{natbib}

\usepackage{graphicx}
\usepackage{color}
\usepackage{subfigure}
\usepackage{amsmath}
\usepackage{amssymb}
\usepackage{mathrsfs}
\usepackage{appendix}
\newcommand{\bk}{\boldsymbol k}
\newcommand{\bh}{\boldsymbol h}
\newcommand{\bA}{\boldsymbol A}
\newcommand{\bC}{\boldsymbol C}
\newcommand{\bM}{\boldsymbol M}
\newcommand{\be}{\boldsymbol e}
\newcommand{\balpha}{\boldsymbol \alpha}
\journalname{Journal of Mathematical Biology}
\def\makeheadbox{\relax} 
\begin{document}

\title{FDG kinetics in cells and tissues: a biochemically-driven compartmental approach
}
\titlerunning{FDG kinetics in cells and tissues}        

\author{Mara Scussolini  \and
	  Vanessa Cossu \and
	  Cecilia Marini \and
	  Gianmario Sambuceti \and
          Giacomo Caviglia 
}

\institute{M. Scussolini \at
              Dipartimento di Matematica, Universit\`a di Genova, Via Dodecaneso 35, 16146 Genova, Italy \\
              \email{scussolini@dima.unige.it}  
           \and
           V. Cossu \at Dipartimento di Medicina Nucleare, IRCCS-IST San Martino, Salita Superiore della Noce 29, 16131 Genova, Italy 
           \and
           C. Marini \at Dipartimento di Medicina Nucleare, IRCCS-IST San Martino, Salita Superiore della Noce 29, 16131 Genova, Italy, 
           and Dipartimento di Scienze della Salute, Universit\`a di Genova, Via Antonio Pastore 1 16132, Genova, Italy, 
           and CNR Istituto di Bioimmagini e Fisiologia Molecolare (IBFM), Via Fratelli Cervi 93, 20090 Milano, Italy 
           \and
           G. Sambuceti \at 
           Dipartimento di Medicina Nucleare, IRCCS-IST San Martino, Salita Superiore della Noce 29, 16131 Genova, Italy, 
           and Dipartimento di Scienze della Salute, Universit\`a di Genova, Via Antonio Pastore 1, 16132 Genova, Italy         
           \and
           G. Caviglia \at
           Dipartimento di Matematica, Universit\`a di Genova, Via Dodecaneso 35, 16146 Genova, Italy
}

\date{}

\maketitle

\begin{abstract}
The radioactive glucose analogue 2-deoxy-2-[$^{18}$F]fluoro-D-glucose (FDG) is widely used to reconstruct glucose metabolism and other biological functions in cells and tissues. 
The analysis of data on the time course of FDG tracer distribution is performed by the use of appropriate compartmental models. Motivated by recent results in cell biochemistry, we describe a new compartmental model aiming at the reconstruction of tracer kinetics in cells and tissues, which emphasizes the different roles of the cytosol and of the endoplasmic reticulum.  
Two applications of the new model are examined, that are concerned with real data from cancer cell cultures \emph{in vitro}, and cancer tissues \emph{in vivo}.
The results are compared with those obtained through application of more standard compartmental models against the same datasets and appear to be in a better agreement with respect to recent biochemical experimental evidence. In particular, it is shown that tracer tends to accumulate in the endoplasmic reticulum, rather than cytosol, and that the rate of phosphorylation is higher than predicted by current models. 


\end{abstract}

\section{Introduction} \label{intro}

The radiopharmaceutical tracer 2-deoxy-2-[$^{18}$F]fluoro-D-glucose (FDG) is extensively used to reconstruct glucose metabolism in cells and tissues, especially in nuclear medicine. 
Following glucose path, FDG is first transported through cell membranes and is then trapped inside cells by phosphorylation. However, unlike phosphorylated glucose, phosphorylated FDG tends to accumulate in cells. For this reason, the measurable radioactive amount of FDG is considered an accurate marker of overall glucose uptake and consumption by cells and tissues \citep{Cherry,Schmidt,Wernick}. In addition, FDG assumption by cancer cells is increased by the Warburg effects for glucose \citep{Vander}; consequently, FDG is used in cancer detection and staging, and to assess the effectiveness of medical treatments.

A basic datum for a detailed analysis of FDG kinetics is the time course of FDG concentration. Concentration of FDG in a suitable region of interest of the target tissue \emph{in vivo} is reconstructed by the use of Positron Emission Tomography (PET) \citep{Bailey,Ollinger}. In a forthcoming work by our group (Scussolini et al. manuscript in preparation), the time dependent activity curve of FDG uptake by a cancer cell culture has been measured also \emph{in vitro} by the use of a LigandTracer (LT) device of Ridgeview Instruments AB Sweden. The LT technology was first described in \cite{Bjorke1} and \cite{Bjorke2}. 

In general, the measured time dependent radioactive signal coming from a target biological system results from superposition of signals emitted by FDG sources occupying, e.g., interstitial tissue, blood, and cells, possibly in either free or phosphorylated forms. Since available measurement devices cannot resolve single emitters, a compartmental model approach is applied, whereby a detailed characterization of tracer kinetics can be reconstructed \citep{Watabe}. 
Essentially, compartments represent uniform spatial distributions or specific chemical compounds of the basic radioactive molecules; radioactivity concentrations in the various compartments are the natural state variables of the system; tracer flow, resulting from interchange of radioactive molecules between compartments, is modeled by a Cauchy problem for a system of linear ordinary differential equations (ODEs) for concentrations; the constant coefficients, also called rate constants or kinetic parameters, represent tracer kinetics and may be related to the action of enzymes, such as hexokinase (HK) responsible for phosphorylation in cells. 
 
In typical compartmental problems the rate coefficients are unknown. 
The measured data are the total amount of tracer (concentration or activity) in a given region of interest, and the input function (IF), describing the time rate of tracer carried into the system. 
Tracer kinetics results from the solution of the inverse problem of determining the unknown rate coefficients compatible with the data, and the subsequent explicit determination of the concentration (or activity) of each compartment through the solution of the system of ODEs.  
In the applications presented in this paper, the inverse problem is solved in two steps: first, a formal expression of the solution of the direct Cauchy problem is evaluated, where dependence on the unknown rate constants is made explicit; second, an inversion algorithm is applied in order to recover the kinetic parameters through comparison of the formal solution with the data. The inversion algorithm makes use of an optimization regularization method which is based on a Newton-type algorithm.

\begin{figure}[htb]
\centering
{\includegraphics[width=7cm]{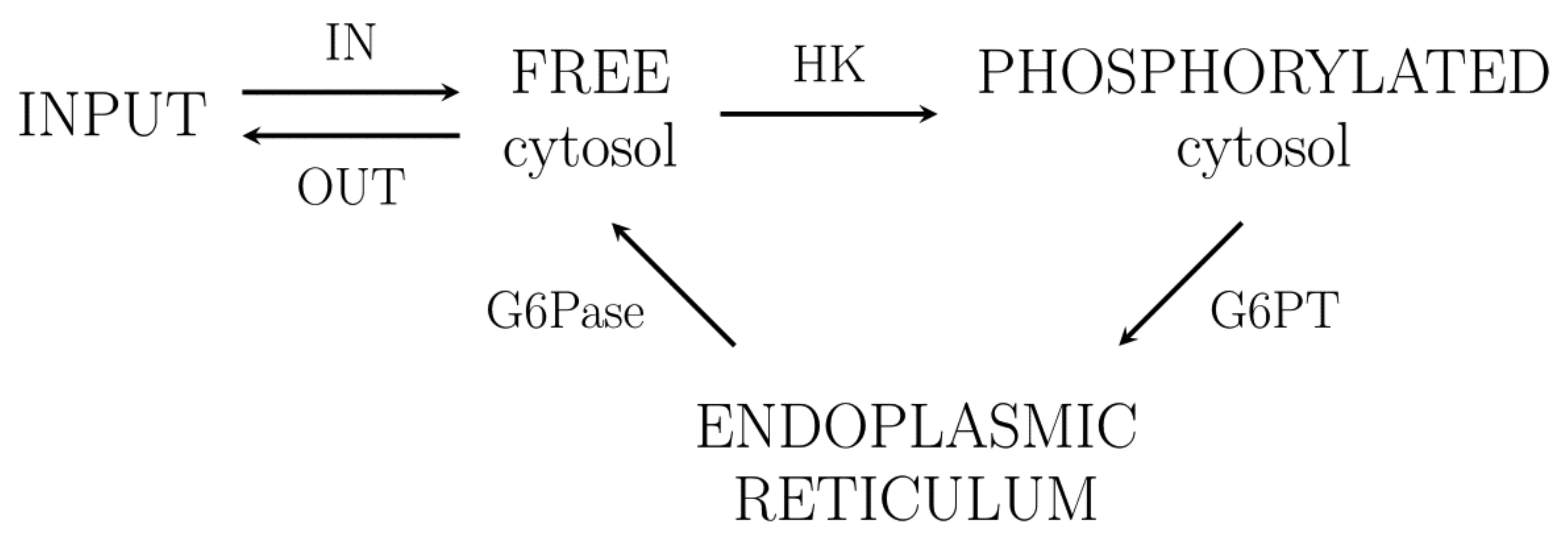}} 
\caption{The biochemical path of FDG inside the cell: the FDG moves in and out the cell environment thanks to GLUT transporters; once inside the cytosol of the cell, free FDG is phosphorylated by hexokinase (HK) and the phosphorylated FDG (FDG6P) can enter the endoplasmic reticulum (ER) transported by G6PT; only inside the ER, FDG6P can be dephosphorylated by G6Pase, after which the FDG turns back in a free status and is released out into the cytosol} 
\label{fig:FDG_kinetics}
\end{figure}

``Classical'' compartmental models have been developed under the assumption that phosphorylation and dephosphorylation of FDG occur in the same intracellular cytosolic volume, as described by \cite{Sokoloff} and \cite{Wernick}. Recent progresses in cell biochemistry have shown that the appropriate location of dephosphorylation is the endoplasmic reticulum (ER) \citep{Ghosh}, which is spatially separated from cytosol. The schematic path of FDG kinetics is illustrated in \figurename~\ref{fig:FDG_kinetics}. 

Following this pattern for FDG kinetics in the cell, here we describe and formalize a new model consisting of three compartments which account for free FDG in cytosol, phosphorylated FDG in cytosol, and phosphorylated FDG in ER. The new biochemically-driven compartmental model is referred to as BCM; a classical simplified compartmental model (SCM) is recovered from the proposed model under the assumption that the ER is removed from consideration.
In the forthcoming paper (Scussolini et al. manuscript in preparation), as a first test of its feasibility, the new BCM has been directly applied to the analysis of data coming from highly controlled \emph{in vitro} experiments on FDG uptake by cell cultures. 
Attention has been concentrated on: (1) the calibration procedure of the LT device, which has been used for the generation of the data; (2) the examination of tracer kinetics when cells were exposed to different glucose concentrations, in order to assess FDG-glucose competition; (3) the biological interpretation of the results. It has been found that tracer tends to accumulates in the ER, a result which has been confirmed by direct measurement on cells seeded \emph{in vitro} and immersed in fluorescent 2DG analogue NBDG; moreover, the value of the rate constant for phosphorylation estimated by application of BCM is greater than that produced by classical SCM, and shows better agreement with results of direct measurements available in the literature \citep{Gao,Muzi}. \\

The analysis of the forthcoming paper by Scussolini et al. (manuscript in preparation) embodies the fundamental role of ER in the description and understanding of tracer kinetics of cancer cell cultures. On this basis, the main aims of the present paper are described as follows.  
(1) To re-examine FDG kinetics in a single cell, in order to construct a new general compartmental model for tracer kinetics capable of being extended to more complex systems, such as cell cultures and tissues, and to verify its applicability. 
(2) To analyze the mathematical properties of BCM, such as identifiability, and the connections with the different types of available data. 
(3) To compare reconstructions of tracer kinetics following from application of BCM and standard SCM to the same set of data, concerning either cancer cell cultures or cancer tissues. 
(4) To discuss consistency and interpretation of the results obtained from applications to cell cultures and tissues. 
(5) To confirm tracer accumulation in ER, and increase in the estimated value of the phosphorylation rate in the new broader framework. 
For completeness and for ease of comparison and interpretation, results on cell cultures are briefly reviewed and new data are analyzed with respect to the forthcoming paper (Scussolini et al. manuscript in preparation). 

The new mathematical models, and the related features, for compartmental analysis of tracer kinetics in cell cultures and tissues are introduced and examined in Section 2. 
Section 3 deals with application on data from cell cultures \emph{in vitro}, while Section 4 deals with data from tissues \emph{in vivo}. Our comments and conclusions on the results are offered in Section 5.

\section{Basic mathematical model for tracer kinetics in cell cultures and tissues} \label{sec:math_model}

FDG kinetics provides an analogue of glucose metabolism in cells and tissues. Starting from local measurements on the diffusion of these radioactive molecules, it allows a quantification of functions in living cells, such as rates of activity of enzymes. 
To this aim, a suitable set of different functional compartments is identified in the assigned target, where each compartment is associated with a specific metabolic state of the tracer, possibly contained in a predefined physiologic volume. 
Tracer flow corresponds to exchange of radioactive molecules between compartments. 
Most considered approaches to tracer kinetics are based on application of such compartmental models. 

In this section we examine a new compartmental model originating from the analysis of the biochemical path of FDG, and partially of glucose, in a single cell. The mathematical counterpart of this compartmental description of intracellular tracer kinetics is given by a system of three linear ODEs for three unknown concentrations and with five unknown constant coefficients. We recall that the constant coefficients are the rate constants. 
Extensions to cell culture and tissue systems are then introduced, illustrating how the general scheme is adapted to the analysis of data provided by allowable measurement devices. In so doing, additional parameters of physiologic interest are introduced, in order to formulate more realistic models. A simplified model is also examined, for the ease of comparison with most diffused existing models. 
To go deeply into the mathematical aspects, we deal with the problem of retrieving the rate constants as solutions of an inverse problem. The corresponding uniqueness problem is also discussed.     

\begin{figure}[htb]
\centering
\subfigure[BCM \label{fig:model_BCM}]
{\includegraphics[width=7cm]{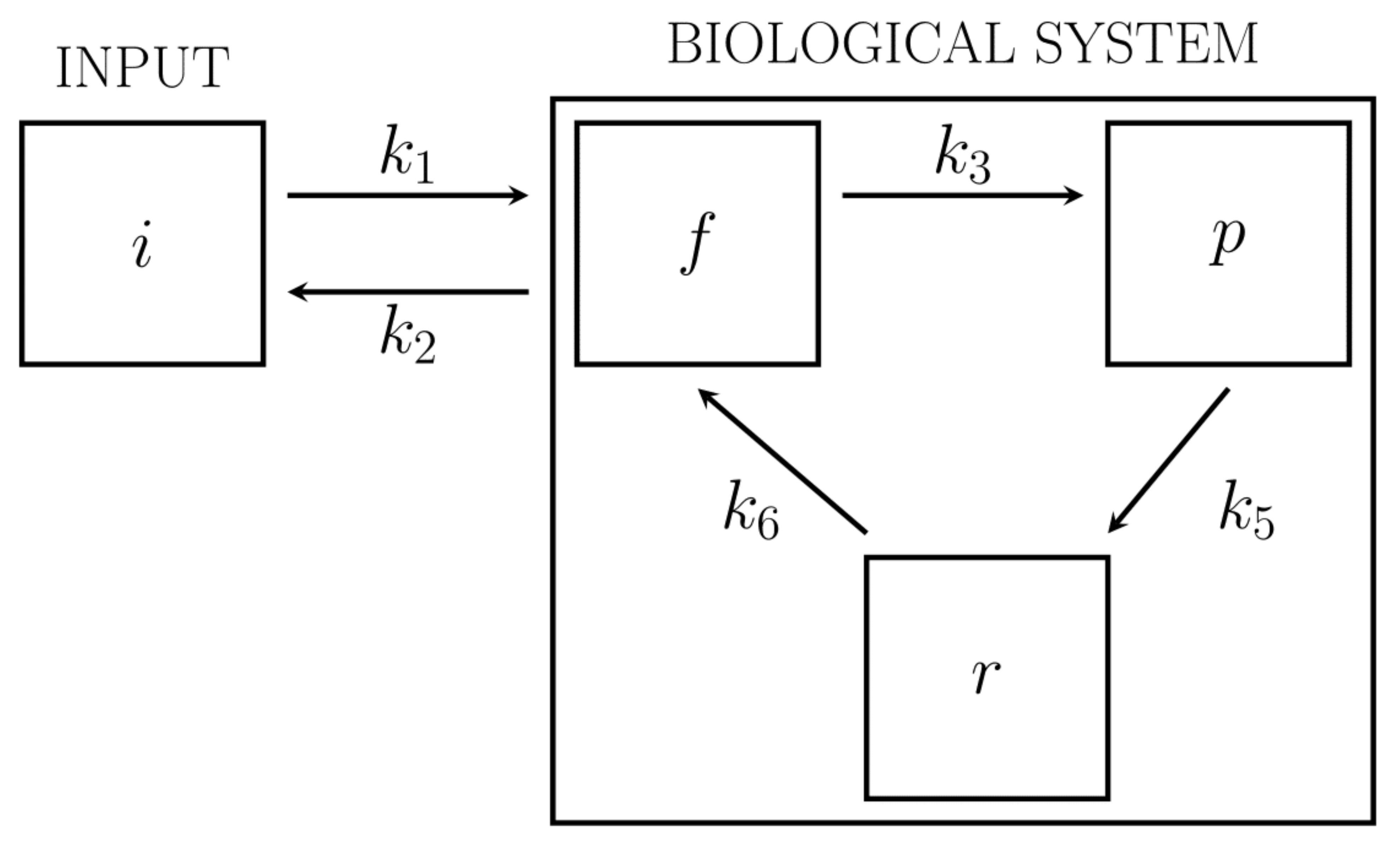}} \\
\subfigure[SCM \label{fig:model_SCM}]
{\includegraphics[width=7cm]{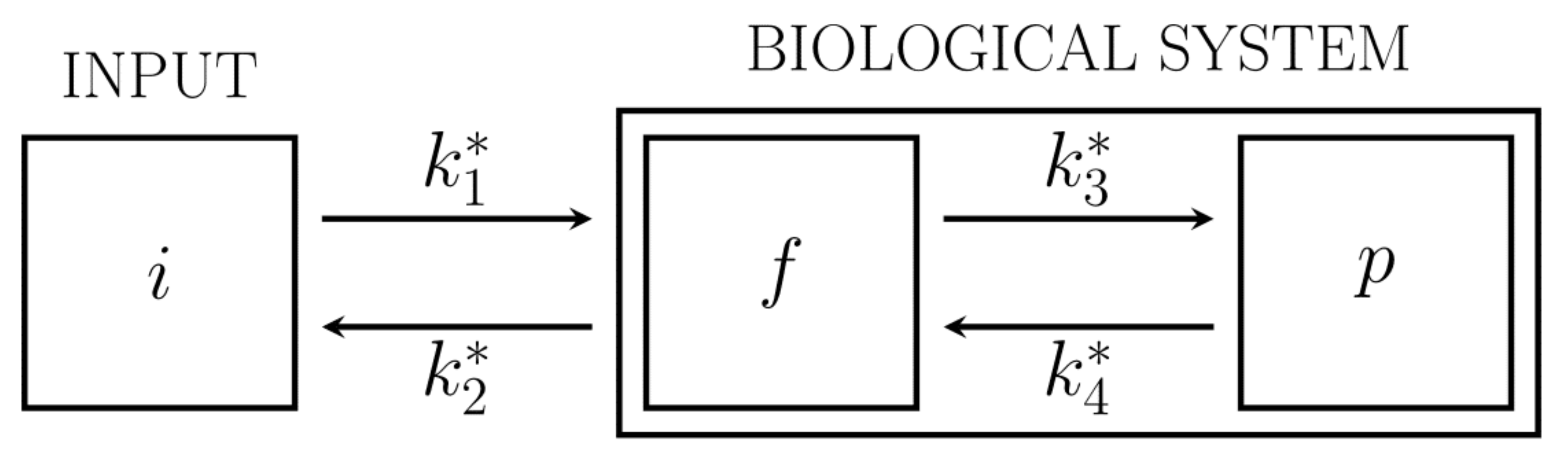}} 
\caption{The two compartmental models considered in this work: (a) the biochemically-driven compartment model (BCM) accounting for compartments $i$ of input tracer, $f$ of free tracer in the cytosol, $p$ of cytosolic phosphorylated tracer, and $r$ of ER-localized phosphorylated tracer. (b) The simplified compartmental model or Sokoloff-type compartmental model (SCM) considering the input pool $i$, the tracer in a free status inside the cell $f$ and the phosphorylated tracer trapped by the cell $p$. The arrows connecting the functional compartments represent the model kinetic parameters, which are denoted as $k$ for the BCM and as $k^*$ for the SCM} 
\label{fig:models}
\end{figure}

\subsection{The model: single cell perspective} \label{subsec:cell_orig}

Consider a cell which is in contact with a liquid containing glucose at physiologic concentration and FDG at a smaller concentration, so that FDG may be regarded as a perturbation of glucose. This general situation is representative of any cells coming into contact with glucose and FDG, both  {\it in vitro} and {\it in vivo}. The biochemical path of glucose and FDG uptake inside the cell may be characterized according to the following scheme. FDG is transported into the cytosol, and back, by glucose transport proteins (GLUT). Inside the cell, glucose and FDG are phosphorylated by hexokinase (HK) to G6P and FDG6P, respectively. Once phosphorylated, glucose continues along the metabolic pathway of glycolysis and pentose-phosphate pathway or participates to glycogen synthesis; instead, FDG cannot follow the same channels and accumulates intracellularly as FDG6P. It is well known that FDG6P is a substrate for G6Pase but, according to recent advances in biological chemistry, G6Pase is anchored to the endoplasmic reticulum (ER) \citep{Ghosh} so that its action of hydrolysis of FDG6P, resulting in the creation of a phosphate group and free tracer, occurs after FDG6P has been transported into the ER lumen by glucose 6-phosphate transporter (G6PT). Subsequently, the free tracer is released into the cytosol. 

The whole process is illustrated symbolically in \figurename~\ref{fig:model_BCM}, which in turn is consistent with \figurename~\ref{fig:FDG_kinetics}. The  ``squares''  $i$, $f$, $p$, $r$ identify the compartments associated with the main steps of tracer kinetics. Specifically, lower indexes $i$, $f$, $p$, and $r$ refer systematically to tracer in the input pool, free tracer in the cytosol, phosphorylated tracer in the cytosol, phosphorylated tracer in the ER. 
In principle, a pool for free tracer in ER could also be considered, which receives tracer also from the free compartment in cytosol; here we assume that its equilibrium value is reached almost instantaneously at the beginning of the experiment and represents a small fraction of tracer contained in ER, so that it is discarded. 

We assume that standard assumptions for application of compartmental models are satisfied. In particular, underlying physiological processes and molecular interactions are not affected by the presence of tracer, distribution of tracer in each compartment is spatially homogeneous, and tracer exchanged between compartments is instantaneously mixed \citep{Cherry,Schmidt,Wernick}.
We also assume that appropriate correction for the physical decay of radioactivity has been applied. 

We denote by $c_f$, $c_p$, and $c_r$ the time dependent and decay corrected concentrations of tracer in the compartments inside a cell, which are regarded as the state variables. 
The concentration of tracer in the external medium, $c_i$, is the given input function of the system. The system of ODEs for the biochemically-driven compartmental model (BCM) is
\begin{equation} \label{eq:dot_c} 
\begin{cases}
\dot{c}_f = - (k_2+k_3) \, c_f + k_6 \, c_r + k_1 \, c_i \\
\dot{c}_p = k_3 \, c_f - k_5 \, c_p  \\
\dot{c}_r = k_5 \, c_p -  k_6 \, c_r   
\end{cases} \ ,
\end{equation}
where the superposed dot denotes the time derivative, and explicit reference to time dependence is omitted. Time is measured in  minutes. The initial conditions are $c_f(0) = c_p(0) = c_r(0) = 0$, which mean that there is no tracer amount in the cell at the beginning of the experiment. The rate constants $k_i$ (1/min), with $i \in \{1,2,3,5,6\}$, describe the first order process of tracer transfer between compartments. In each equation, products of rate constants and concentrations represent fluxes of tracer per unit time and unit volume; plus and minus signs refer to incoming and outgoing fluxes, respectively. 
The system (\ref{eq:dot_c}) expresses conservation of the tracer interchanged between free, phosphorylated and reticular compartments. In view of natural applications and for the ease of comparison of the results, we have adopted the usual notations of nuclear medicine for the rate constants, as done by \cite{Cherry} and by \cite{Wernick}.

Consistently with \figurename~\ref{fig:model_BCM}, the parameters $k_1$ and $k_2$ are the kinetic parameters for transport of FDG from medium to cell and back from cell to medium, respectively; $k_3$ is the phosphorylation rate of FDG;  $k_5$ is the input rate of FDG6P into ER; $k_6$ refers to the dephosphorylation rate of FDG6P to FDG. 
Since the dephosphorylation occurs only inside the ER, a parameter $k_4$, corresponding to an arrow from $p$ to $f$, is not considered. 
The values of the rate constants depend on the conditions of the experiment: for example, they are influenced by the amount of glucose present in the surrounding environment. 
 
Consider the system (\ref{eq:dot_c}) with vanishing initial conditions and  define the vector $\bk_5 = (k_1, k_2, k_3, k_5, k_6)$. The \emph{direct problem} consists in finding the solution of the system of ODEs in the unknowns $c_f$, $c_p$, and $c_r$, for a given vector $\bk_5$ and input function $c_i$. Here, we are mainly concerned with the \emph{inverse problem} of finding FDG kinetics, i.e., determining the vector $\bk_5$ of rate coefficients which corresponds to a given set of data. Following standard approaches, the data are the input function $c_i$ and the total tracer concentration $c_T$ inside the cell. 
The connection between $c_T$ and the state variables is obtained as follows.

Denote by $v_\text{cyt}$ and $v_\text{er}$ the cell volumes of cytosol and ER, respectively, where intracellular tracer is located. The total activity $a_T$ inside the cell is given by 
\begin{equation} \label{eq:a_T} 
a_T = v_\text{cyt} \, (c_f+c_p) + v_\text{er} c_r \ . 
\end{equation}
On letting $c_T = a_T / (v_\text{cyt}+v_\text{er}) $ be the total density, it is found that
\begin{equation} \label{eq:c_T} 
c_T = (1-v_r) \,(c_f+c_p) + v_r \,c_r \ ,
\end{equation}
where $v_r =  v_\text{er} / (v_\text{cyt}+ v_\text{er})$ is the volume fraction of the ER. 
Henceforth we regard $v_r$ as a given experimental parameter. The expression 
(\ref{eq:c_T}) for the total density agrees with similar expression that can be found in the literature on tracer kinetics in tissues (see, e.g., \cite{Wernick}). 
 
In principle, $c_T$ (or $a_T$) may be regarded as a physical quantity whose time course is determined by measurement procedures. Then (\ref{eq:c_T})  provides the connection between the measured quantity $c_T$ and the solution of the system (\ref{eq:dot_c}), expressed in terms of $\bk_5$. Therefore, (\ref{eq:c_T}) is the starting point for the solution of the inverse problem. 
 
\subsection{Cell cultures and tissue systems} \label{subsec:conc_C}  

The crucial point of the previous formulation is the remark that, usually, a single cell is not accessible to measurements of radiation emitted in time. Applications are based on observation of the time course of radiation emitted by cell cultures \emph{in vitro} or tissues \emph{in vivo}. Thus the model described in \figurename~\ref{fig:model_BCM} has to be adapted to applications on a higher scale. For example, if we consider a colony of $N$ cells, it is natural to define the concentration of free tracer of the colony as $C_f = N\, c_f$, and so on. Similarly, we denote by $C_f$ the concentration of free tracer in a given tissue. 
 
Accordingly, we introduce the (macroscopic) state variables $C_f$, $C_p$, and $C_r$ which describe the concentrations of free tracer, phosphorylated tracer in cytosol, and phosphorylated tracer ER, respectively. It is assumed that the corresponding system of ODEs takes the form (\ref{eq:dot_c}), which is rewritten compactly as
 \begin{equation} \label{eq:ode_C} 
\dot{\bC}= \bM \, \bC + k_1 \, C_i \, \be \qquad \bC(0)= \mathbf{0} \ ,
\end{equation}
where
\begin{equation} \label{eq:sode_def_matrici_4} 
\bM = \begin{bmatrix} 
- (k_2+k_3) & 0 & k_6  \\
k_3 & -k_5 & 0  \\ 
0 & k_5 & -k_6 \end{bmatrix} \, , \ 
\bC = \begin{bmatrix}  C_f \\ C_p \\ C_r    \end{bmatrix} \, , \  
\be = \begin{bmatrix}  1 \\ 0 \\ 0  \end{bmatrix} \ ,
\end{equation} 
and where $C_i$ is the given input function. 
The rate constants maintain the interpretation discussed in the previous subsection.
The analytic solution of the Cauchy problem (\ref{eq:ode_C}) takes the form 
\begin{equation} \label{eq:sol_C} 
\bC(t;\bk_5,C_i) = k_1 \, \int_0^t  e^{\bM \,(t-\tau) } \, \be \, C_i(\tau)   \,d\tau \ , 
\end{equation} 
with the time variable $t \in \mathbb R_+$.

The system (\ref{eq:ode_C}) is regarded as the basic mathematical formulation of compartmental analysis adopted in this work. Equation (\ref{eq:sol_C}) provides the solution of the direct problem, at given $\bk_5$. Before dealing with the inverse problem, the connection with data for both cell cultures and tissues must be considered rather carefully.   

\subsubsection{Data for cell cultures and activity formulation} \label{subsubsec:data_cells}

Available data on the time course of radioactivity for a cell culture are given in terms of total activity. In principle, rephrasing of the data in concentrations was allowed but this required, at least, the knowledge of parameters such as the total number of cells, and the volumes of cytosol and ER. These parameters are only roughly known. Besides other advantages, the reformulation of the system (\ref{eq:ode_C}) in activities allows to reduce the parameters to one, precisely, the ratio between the volumes of the cytosol and ER, which can be estimated and is independent of the number of cells. \\

Concentrations and corresponding activities of the cell culture are related by 
\begin{equation} \label{eq:conc_act_0}
C_f = \frac{A_f}{\mathscr{V}_\text{cyt}} \, , \qquad  C_p = \frac{A_p}{\mathscr{V}_\text{cyt}} \, , \qquad C_r = \frac{A_r}{\mathscr{V}_\text{er}} \, , \qquad  C_i = \frac{A_i}{\mathscr{V}_\text{i}} \, ,
\end{equation} 
where $\mathscr{V}_\text{cyt}$, $\mathscr{V}_\text{er}$, and $\mathscr{V}_\text{i}$ are the volumes of the total cytosolic region, ER, and external liquid, respectively. 
Substitution of the activities into the system (\ref{eq:ode_C}) leads to the formulation of the Cauchy problem  
 \begin{equation} \label{eq:ode_A} 
\dot{\bA}= \bM \, \bA + \bar{k}_1 \, A_i \be \ , \qquad \bA(0)= \mathbf{0} \ ,
\end{equation}
where
\begin{equation} \label{eq:sode_def_matrici_A} 
\bA = \begin{bmatrix} A_f \\ A_p \\ \bar{A}_r \end{bmatrix} \ , \qquad  
\bar{A}_r = A_r \, \frac{\mathscr{V}_\text{cyt}}{\mathscr{V}_\text{er}} \ , \qquad \bar{k}_1 = k_1 \, \frac{\mathscr{V}_\text{cyt}}{\mathscr{V}_\text{i}} \ ,
\end{equation} 
and where $A_i$ is the given input function, representing in our case the total activity of the incubation medium in which cells are immersed. 
Notice that the matrices $\bM$ and $\be$ are left unchanged by the transformation of the state variables.  
The auxiliary variable $\bar{A}_r$ is related  to the ``natural'' activity $A_r$ of the ER through the dimensionless ratio $\mathscr{V}_\text{cyt}/\mathscr{V}_\text{er}$, which is independent of the number of cells and coincides with $v_\text{cyt}/v_\text{er}$. Accordingly, we find 
\begin{equation} \label{eq:A_bar} 
A_r = v \, \bar{A}_r \ , \qquad v = \frac {\mathscr{V}_\text{er}} {\mathscr{V}_\text{cyt}} =
\frac {v_\text{er} } {v_\text{cyt} } \ ,
\end{equation} 
where in particular $v < 1$ \citep{Milo}.
The coefficient $\bar{k}_1 = k_1 \mathscr{V}_\text{cyt}/\mathscr{V}_\text{i}$ provides the rate constant adapted to the description in terms of activities and plays the same role as $k_1$. The other coefficients $k_2$, $k_3$, $k_5$, and $k_6$ preserve the interpretation as rate constants and the numerical value pertaining to the system (\ref{eq:ode_C}). 

The analytic solution of the Cauchy problem (\ref{eq:ode_A}) takes the form 
\begin{equation} \label{eq:sol_A} 
\bA(t;\bar{\bk}_5,A_i) = \bar{k}_1 \, \int_0^t  e^{\bM \,(t-\tau) } \, \be \, A_i(\tau)   \,d\tau \ ,
\end{equation} 
where, by a slight abuse of language, we let $\bar{\bk}_5=(\bar{k}_1,k_2,k_3,k_5,k_6)$.

Denote as $\mathcal{A}_T$ the measured time course of the total activity of the cell culture. Following the analogy with equation (\ref{eq:a_T}), we have
\begin{equation} \label{eq:master_eq_act_0}
\mathcal{A}_T = A_f + A_p + A_r = A_f + A_p + v \,\bar{A}_r \ .
\end{equation}
Equation (\ref{eq:master_eq_act_0}) may be written in compact form as
\begin{equation} \label{eq:master_eq_act_1}
\mathcal{A}_T(t) = \balpha \, \bA(t;\bar{\bk}_5,A_i) \ , \qquad \balpha = \begin{bmatrix} 1 & 1 & v \end{bmatrix} \ .
\end{equation}
Equation (\ref {eq:master_eq_act_1}) is the basic equation for the formulation of the inverse problem 
of determining the rate coefficients, in that it relates measured quantities of cell culture system to formal expressions of the unknown vector $\bar{\bk}_5$.

\subsubsection{Data for tissues} \label{subsubsec:data_tissues}

Widely applied models for tracer kinetics in tissues assume that tracer is initially injected into blood. Next it is carried from blood to tissues and cells; once it has reached the target tissue, it may remain in cells, mainly in phosphorylated form, or may be transported back to blood in free form. A small percentage of phosphorylated tracer can be dephosphorylated. In the corresponding compartmental formulation it has been customary to consider a compartment corresponding to blood, and two tissue compartments, for free and phosphorylated tracer, respectively. In particular, the compartment for free tracer accounted for both free interstitial tracer and free intracellular tracer \citep{Schmidt,Sokoloff}. 

In order to insert into the tissue scheme details on tracer kinetics in cells, we consider:
\begin{itemize}
\item a blood compartment of concentration $C_i$, providing the input function;
\item a compartment of concentration $C_f$, for free tracer in the interstitial space and in the cytosol of tissue cells;
\item a compartment of concentration $C_p$, for phosphorylated tracer in cytosol;
\item a compartment of concentration $C_r$, for phosphorylated tracer in ER.
\end{itemize}
According to this BCM approach, tracer kinetics is still described by the Cauchy problem (\ref{eq:ode_C}) for concentrations, with analytic solution $\bC$ given by equation (\ref{eq:sol_C}). 

The data are the input function $C_i$ and the concentration $\mathcal{C}_T$, measured over a suitable region of interest, belonging to the target tissue.   
We show that $\mathcal{C}_T$ is a weighted sum of the state variables $C_f$, $C_p$, $C_r$, and $C_i$.

The volume $\mathscr{V}_\text{tot}$ of the region of interest may be partitioned as
\begin{equation} \label{eq:V_tot}
\mathscr{V}_\text{tot} = \mathscr{V}_\text{blood} + \mathscr{V}_\text{int} + \mathscr{V}_\text{cyt} + \mathscr{V}_\text{er} \ ,
\end{equation} 
where $\mathscr{V}_\text{blood}$ and $\mathscr{V}_\text{int}$ denote the volume occupied by blood and interstitial fluid, respectively; extending previous notations, $\mathscr{V}_\text{cyt}$ and $\mathscr{V}_\text{er}$ denote total volumes of cytosol and ER of the tissue cells.
The total activity $\mathcal{A}_T = \mathscr{V}_\text{tot} \, \mathcal{C}_T$ of the tracer occupying the volume of interest is related to the state variables and the input function by the equation
\begin{equation} \label{eq:A_T}
\mathscr{V}_\text{tot} \, \mathcal{C}_T = \mathscr{V}_\text{blood} \,C_i + \mathscr{V}_\text{int} \,C_f+ \mathscr{V}_\text{cyt}\,C_f + \mathscr{V}_\text{cyt}\,C_p + \mathscr{V}_\text{er} \,C_r \ .
\end{equation} 
Division of both sides by $\mathscr{V}_\text{tot}$ leads to
\begin{equation} \label{eq:C_T_tiss}
\mathcal{C}_T = \frac { \mathscr{V}_\text{blood} } {\mathscr{V}_\text{tot}} \,C_i + 
\frac{\mathscr{V}_\text{int}+\mathscr{V}_\text{cyt}} {\mathscr{V}_\text{tot}} \,C_f+ \frac{ \mathscr{V}_\text{cyt}} {\mathscr{V}_\text{tot}} \, C_p + \frac {\mathscr{V}_\text{er} }{\mathscr{V}_\text{tot}} \,C_r \ .
\end{equation} 
We define the volume fractions of blood and interstitial fluid as 
\begin{equation} \label{eq:V_b_V_i}
V_b = \frac { \mathscr{V}_\text{blood} } {\mathscr{V}_\text{tot}} \ , \qquad 
V_i = \frac {\mathscr{V}_\text{int}} {\mathscr{V}_\text{tot}} \ .
\end{equation} 
Next application of (\ref{eq:V_tot}) and (\ref{eq:V_b_V_i}) yields
\begin{equation} \label{eq:frac_er}
\frac {\mathscr{V}_\text{er} }{\mathscr{V}_\text{tot}} = v_r \, (1-V_b-V_i) \ ,
\end{equation}
where 
\begin{equation} \label{eq:v_r}
v_r = \frac {\mathscr{V}_\text{er} }{\mathscr{V}_\text{cyt} + \mathscr{V}_\text{er}} = 
 \frac {v_\text{er} }{v_\text{cyt} + v_\text{er}} 
\end{equation}
is independent of the number of cells. 
Similarly, comparison with (\ref{eq:V_tot}),  (\ref{eq:V_b_V_i}), and (\ref{eq:frac_er}) provides
\begin{equation} \label{eq:frac_cyt}
\frac {\mathscr{V}_\text{cyt} }{\mathscr{V}_\text{tot}} =(1 -v_r) \, (1-V_b-V_i) \ .
\end{equation}
Replacing (\ref{eq:V_b_V_i}), (\ref{eq:frac_er}), and (\ref{eq:frac_cyt}) into the expression (\ref{eq:C_T_tiss}) of $\mathcal{C}_T$ provides the required result
\begin{equation} \label{eq:C_T_tiss_pesato}
\mathcal{C}_T = V_b \,C_i + \alpha \, C_f + \beta \,C_p + \gamma \,C_r \ ,
\end{equation}
where the adimensional constants $\alpha$, $\beta$ and $\gamma$ are defined as
\begin{align}
\label{eq:alpha}
 \alpha & = V_i + (1 -v_r) \, (1-V_b-V_i), \\  
\label{eq:beta} 
\beta  & = (1 -v_r) \, (1-V_b-V_i)  , \\
\label{eq:gamma} 
\gamma & = v_r \, (1-V_b-V_i) .
\end{align}
In compact form, we can write equation (\ref{eq:C_T_tiss_pesato}) as
\begin{equation} \label{eq:C_T_tiss_pesato_comp}
\mathcal{C}_T(t) = V_b \,C_i(t) + \balpha \, \bC(t;\bk_5, C_i) \ , \qquad \balpha = \begin{bmatrix} \alpha & \beta & \gamma \end{bmatrix} \ , 
\end{equation}
where $V_b$ and $\balpha$ depend on specific tissue and cell features; they are regarded as given in the inversion procedure. 

\subsection{Simplified BCM} \label{subsec:SCM}

A simplified compartmental model (SCM) for compartmental analysis is shown in \figurename~\ref{fig:model_SCM}. As in the compartmental system of  \figurename~\ref{fig:model_BCM}, tracer is first exchanged between the input compartment (either incubation medium or blood) and the compartment for free tracer, with rate coefficients $k_1^*$ and $k_2^*$. Unlike the system of \figurename~\ref{fig:model_BCM}, there is only one cytosolic pool for phosphorylated FDG. The coefficients $k_3^*$ and $k_4^*$, providing phosphorylation and dephosphorylation rates, can be regarded as the functional correspondent of $k_3$ and $k_6$, respectively. The model of \figurename~\ref{fig:model_BCM} is known conventionally as the Sokoloff model, first introduced in \cite{Sokoloff}. In view of further comparison, a few details of the SCM formulation are now outlined. 

Following the conventions already introduced, the system of ODEs for the two state variables $C_f^*$ and $C_p^*$ takes the form
\begin{equation} \label{eq:dot_C_S} 
\begin{cases}
\dot{C}_f^* = - (k_2^*+k_3^*) \, C_f^*  + k_1^* \, C_i , \\
\dot{C}_p^* = k_3^* \, C_f^* - k_4 \, C_p^* ,
\end{cases} \ ,
\end{equation}
with initial conditions $C_f^*(0)= C_p^*(0)=0$, and given input function $C_i$. We denote by $\bk_4^* = (k_1^*,k_2^*,k_3^*,k_4^*)$ the vector of parameters of the simplified formulation. Notice that a star is used systematically to refer to quantities pertaining to the simplified model. 

In the modeling of a cell culture, the state variable $C_i$ describes concentration of tracer in the incubation medium, while $C_f^*$ and $C_p^*$ describe intracellular concentration of free and phosphorylated tracer. We do not go into the details of the compact formulation in terms of activities, which is obtained straightforwardly. We only observe that the connection between the datum and the state variables takes the simplified form
\begin{equation} \label{eq:master_eq_act_*} 
\mathcal{A}_T = A_f^*+ A_p^* = \balpha \, \bA^* \ , 
\end{equation}
where
\begin{equation}
\bA^* = \begin{bmatrix} A_f ^* \\ A_p^* \end{bmatrix} \ , \qquad \balpha = \begin{bmatrix} 1 & 1 \end{bmatrix} \ .
\end{equation}
Here, $\bA^* = \bA^*(t;\bar{\bk}_4^*,A_i)$, where $\bar{\bk}_4^* = (\bar{k}_1^*,k_2^*,k_3^*,k_4^*)$ includes the modified parameter $\bar{k}_1$ defined in (\ref{eq:sode_def_matrici_A}). 

In the analysis of a tissue model, $C_i$ corresponds to tracer concentration in blood. The state variable $C_f^*$ is interpreted as the concentration of free tracer in interstitial tissue and tissue cells, while $C_p^*$ is the concentration of phosphorylated tracer inside tissue cells. It is assumed that extracellular phosphorylated tracer can be disregarded. In this simplified framework, the equation connecting data to state variables is
\begin{equation} \label{eq:C_T_tiss_pesato_Sok}
\mathcal{C}_T = V_b \,C_i + (1-V_b) \, C_f^* + (1-V_b-V_i)  \,C_p^* = V_b \,C_i + \balpha \, \bC^* \ , 
\end{equation} 
where  
\begin{equation} 
\bC^* = \begin{bmatrix} C_f ^* \\ C_p^* \end{bmatrix} \ , \qquad \balpha = \begin{bmatrix} 1-V_b, & 1-V_b-V_i \end{bmatrix} \ .
\end{equation} 
Here, $\bC^* = \bC^*(t;\bk_4^*,C_i)$, where $\bk_4^* = (k_1^*,k_2^*,k_3^*,k_4^*)$. 

The SCM can be derived from the BCM simply by omitting consideration of the role of the ER, i.e., by formal substitution of the condition $\mathcal{V}_{er} = 0$. 
If in addition we assume that $V_i=0$, then equation (\ref{eq:C_T_tiss_pesato_Sok}) reduces to 
\begin{equation} \label{eq:C_T_Sok_vero}
\mathcal{C}_T = V_b \,C_i + (1-V_b) \, (C_f^* + C_p^*) \ , 
\end{equation} 
which is the standard equation often used in tissue kinetics. 

In this work the SCM is considered explicitly for the ease of comparison. Specifically, we will apply the BCM and the SCM to the analysis of the same data in order to discuss similarities and differences between the results. 


\subsection{A general relation between rate constants of BCM and SCM} \label{subsec:BCM_vs_SCM}

This subsection is devoted to the determination of a general relation between the rate constants of the BCM and the corresponding SCM, which holds if the two models are consistent with the same data. Under suitable assumptions, this relation is further reduced to a remarkable difference in the rates of phosphorylation $k_3$ and $k_3^*$, which is to be regarded as a direct consequence of the modeling assumptions. 

With the aim of performing a qualitative analysis on the two models BCM and SCM, the following considerations are made.
It is well known that the dephosphorylation rate is rather small \citep{Sokoloff}; therefore, we assume that $k_6$ and $k_4^*$ are small with respect to the other coefficients, so that their contribution can be neglected.
Next, we suppose that the concentrations $C_i$, $C_f$, $C_p$, and $C^*_f$ are almost constant at large time values.
The systems of ODEs (\ref{eq:ode_C}) and (\ref{eq:dot_C_S}) reduce to the algebraic conditions 
\begin{align*}
(k_2+k_3) \, \tilde{C}_f = k_1 \, \tilde{C}_i \\
k_3 \, \tilde{C}_f = k_5 \, \tilde{C}_p  \\
\dot{C}_r= k_5 \, \tilde{C}_p  
\end{align*} 
and
\begin{align*}
 (k_2^*+k_3^*) \, \tilde{C}_f^* =  k_1^* \, \tilde{C}_i \\
\dot{C}_p^* = k_3^* \, \tilde{C}_f^*   
\end{align*}
where the superposed tilde refers to the constant values of the concentrations and $\tilde{C}_i$ is the common forcing contribution, independent of the model. The constant rates of growth of phosphorylated FDG are given by 
\begin{equation} \label{eq:rate_Cr}
\dot{C}_r =  \frac {k_1\, k_3 }{k_2+k_3} \,  \tilde{C}_i \ ,
\end{equation}
\begin{equation} \label{eq:rate_CP}
\dot{C}_p^* =  \frac {k_1^* \,  k_3^*}{k_2^*+k_3^*} \,  \tilde{C}_i \ .
\end{equation} 

Consider the case of cell cultures. Comparison of eqs (\ref{eq:master_eq_act_0}) and (\ref{eq:master_eq_act_*}) for the total activity shows that 
\begin{equation*} 
\mathcal{A}_T = A_f+A_p+A_r = A_f^*+A_p^* \ .
\end{equation*}
In view of the assumptions, evaluation of the time derivative of the last equation leads to $\dot{A}_r= \dot{A}_p^*$, which is written in the equivalent form
\begin{equation} \label{eq:legame_cells}
 \mathscr{V}_\text{er} \,\dot{C}_r =\mathscr{V}_\text{cyt} \dot{C}_p^* \ ,
\end{equation}
after comparison with (\ref{eq:conc_act_0}). Substitution into (\ref{eq:legame_cells}) of (\ref{eq:rate_Cr}),  (\ref{eq:rate_CP}), and the definition (\ref{eq:sode_def_matrici_A}) of $\bar{k}_1$, shows that
\begin{equation} \label{eq:id_k_cells}
\frac {\bar{k}_1^* \,  k_3^*}{k_2^*+k_3^*} = v \,  \frac {\bar{k}_1\, k_3 }{k_2+k_3} \ ,
\end{equation}
where we recall that $v= \mathscr{V}_\text{er}/ \mathscr{V}_\text{cyt}$.
Equation (\ref{eq:id_k_cells}) may be used as a check on the effectiveness of the numerical reconstructions. 

If  $\bar{k}_1^* \approx \bar{k}_1$, $k_2^* \approx k_2$, $k_3^* \ll k_2^*$, and $k_3 \ll k_2$, as it is shown to be the case in subsequent developments, then equation (\ref{eq:id_k_cells}) simplifies to $k_3^* \approx v \, k_3$. This shows that the factor $v$ connects the reconstructed phosphorylation rates of SCM and BCM. 

Similar considerations hold for the case of tissues. Comparison of the expressions (\ref{eq:C_T_tiss_pesato}) and (\ref{eq:C_T_tiss_pesato_Sok}) of the total concentration for the BCM and SCM shows that
\begin{equation*} 
\alpha \, C_f + \beta \,C_p + \gamma \,C_r= (1-V_b) \, C_f^* + (1-V_b-V_i)  \,C_p^* \ .
\end{equation*} 
In view of the original assumptions and the definition of  $\gamma$, evaluation of the time derivative yields
\begin{equation} \label{eq:legame_tissues}
v_r \, \dot{C}_r =  \,\dot{C}_p^* \ .
\end{equation} 
Substitution into (\ref{eq:legame_tissues}) of the expressions (\ref{eq:rate_Cr}) and  (\ref{eq:rate_CP}) leads to equation
 \begin{equation} \label{eq:id_k_tissues}
\frac{ k_1^* \,  k_3^*}{k_2^*+k_3^*} = v_r \,  \frac {k_1\, k_3 }{k_2+k_3} \ ,
\end{equation}
which is similar to (\ref{eq:id_k_cells}), with $v_r$ replacing $v$. 

\subsection{Compartmental inverse problem} \label{subsec:solution_IP}

The compartmental inverse problem consists in finding the rate coefficients of the model, starting from the available data. In this subsection we discuss the two main issues related to the inverse problem: 
the identifiability of the model, assessing whether the parameters are uniquely determined by the given data, and the numerical method applied in order to reduce the compartmental model and return the numerical values of the kinetic parameters. 

\subsubsection{Identifiability issues} \label{subsubsec:id}

Before proceeding to numerical evaluation of the rate coefficients, we discuss the formal identifiability of the model, namely, whether the rate coefficients are uniquely determined by the given input data, under the assumption that they are not contaminated by noise \citep{Miao,Yates}. 
The proof of uniqueness may be regarded as an a priori test on the compartmental  model, assuring that it is effective in providing a unique description of tracer kinetics, independently of the numerical values of the data. We show that the BCM is identifiable for both the tissue and cell culture systems, under general conditions. Identifiability of the tissue model, the more complicated system, is considered first; then the cell culture model is examined. Notice that, it is already well known that the SCM is identifiable, and we refer to \cite{Delbary_id} for the proof. \\

The discussion of identifiability of BCM tissue model is based on the system of ODEs (\ref{eq:ode_C}) and equation (\ref{eq:C_T_tiss_pesato}), with $\mathcal{C}_T$ and $C_i$ given.
Identifiability corresponds to uniqueness of the vector $\bk_5$. Following the procedure
used in \cite{Delbary_id}, we consider the Laplace transform of the system (\ref{eq:ode_C}) and equation (\ref{eq:C_T_tiss_pesato}), in order to reduce the identifiability issue to the proof of uniqueness of the solution of an algebraic system.

We denote by $\tilde{f}(s)$ the Laplace transform of a function $f(t)$. Assuming that suitable regularity conditions are satisfied, we obtain the linear system
\begin{equation} \label{eq:Tissue_C} 
\begin{cases}
(s+k_2+k_3) \, \tilde{C}_f - k_6 \, \tilde{C}_r  = k_1 \, \tilde{C}_i   \\
- k_3 \, \tilde{C}_f + (s+ k_5) \, \tilde{C}_p  = 0 \\
- k_5 \, \tilde{C}_p + (s+ k_6) \, \tilde{C}_r  = 0 
\end{cases}
\end{equation}
for the transform of the system (\ref{eq:ode_C}), and equation
\begin{equation} \label{eq:Tissue_C_T}
\tilde{\mathcal{C}}_T -V_b \, \tilde{C}_i = \alpha\, \tilde{C}_f  + \beta \, \tilde{C}_p   + \gamma \, \tilde{C}_r 
\end{equation}  
from the transform of (\ref{eq:C_T_tiss_pesato}). 

The solution of the linear system (\ref{eq:Tissue_C}) is 
\begin{align}  
\label{eq:til_Cf}
\tilde{C}_f  &= \frac {k_1}{D(s)} \, (s+k_5) \, (s+k_6) \, \tilde{C}_i  \\
\label{eq:til_Cp}
\tilde{C}_p &= \frac {k_1}{D(s)} \, k_3 \, (s+k_6) \, \tilde{C}_i  \\
\label{eq:til_Cr}
\tilde{C}_r &= \frac {k_1}{D(s)} \, k_3 \, k_5 \, \tilde{C}_i  
\end{align}
where 
\begin{equation} \label{eq:D_espl_4C}
D(s) = s^3 + (k_2+k_3+k_5+k_6) \, s^2 + [ (k_2+k_3) \, (k_5+k_6) + k_5 \, k_6 ] \, s + k_2 \, k_5 \, k_6  \ .
\end{equation}
Substitution of the expressions (\ref{eq:til_Cf}), (\ref{eq:til_Cp}), and (\ref{eq:til_Cr}) of $\tilde{C}_f$, $\tilde{C}_p$, and $\tilde{C}_r$ into equation (\ref{eq:Tissue_C_T}) yields the necessary condition
\begin{equation} \label{eq:Tissue_bis_4C}
\frac{\tilde{\mathcal{C}}_T - V_b \, \tilde{C}_i }{ \tilde{C}_i } =  \frac{k_1 \, Q(s)}{D(s)} \ ,
\end{equation}
where
\begin{equation}
Q(s) = \alpha \, s^2 + [\alpha \, ( k_5 + k_6)+ \beta \,k_3]  \, s + \alpha \, k_5 \,k_6  +\beta \, k_3\,k_6 + \gamma \, k_3 \, k_5 \ .
\end{equation} 

If $\bh_5 = (h_1,h_2,h_3,h_5,h_6) $ is another vector of rate coefficients consistent with the data, we have to prove that $\bh_5=\bk_5$. Compatibility with data implies equality between the right-hand sides of (\ref{eq:Tissue_bis_4C}), expressed in terms of $\bh_5$ and in terms of $\bk_5$. With obvious meaning of symbols, we have
\begin{equation} \label{eq:poly_n}
\frac{h_1 \, Q_{\bh_5}(s)}{D_{\bh_5}(s)} = \frac{k_1 \, Q_{\bk_5}(s)}{D_{\bk_5}(s)} \ .
\end{equation}
Assume that the polynomials $Q$ and $D$ are coprime, i.e. they do not have common roots. 
Since the leading coefficients of $ Q_{\bh_5}$ and $Q_{\bk_5}$ are identical, as well as those of $D_{\bh_5}$ and $D_{\bk_5}$, equation (\ref{eq:poly_n}) holds if and only if  $h_1=k_1$,   $D_{\bh_5} = D_{\bk_5}$, and  $Q_{\bh_5} = Q_{\bk_5}$. The last two equations give rise to the system
\begin{equation} \label{eq:sis_1}
 h_2+h_3+h_5+h_6=k_2+k_3+k_5+k_6
\end{equation}
\begin{equation} \label{eq:sis_2} 
(h_2+h_3) \, (h_5+h_6) + h_5 \, h_6 = (k_2+k_3) \, (k_5+k_6) + k_5 \, k_6 
\end{equation}
\begin{equation} \label{eq:sis_3}
h_2 \, h_5 \, h_6 = k_2 \, k_5 \, k_6  
\end{equation}
\begin{equation} \label{eq:sis_4}
\beta\, h_3 + \alpha \,(h_5 + h_6)= \beta\, k_3 + \alpha \,(k_5 + k_6) 
\end{equation}
\begin{equation} \label{eq:sis_5}
 \alpha \, h_5 \,h_6  +\beta \, h_3\,h_6 + \gamma \, h_3 \, h_5  =   \alpha \, k_5 \,k_6  +\beta \, k_3\,k_6 + \gamma \, k_3 \, k_5
\end{equation}
of five equations for the four unknowns $h_2$, $h_3$, $h_5$, $h_6$. 

The analysis of the system (\ref{eq:sis_1})--(\ref{eq:sis_5}) proceeds in three steps. First, eqs (\ref{eq:sis_1}), (\ref{eq:sis_3}), (\ref{eq:sis_4}) are solved for $h_3$, $h_5$ and $h_6$ in terms of $h_2$. Next equation (\ref{eq:sis_2}) is solved for $h_2$ in terms of $\bk_5$, $\alpha$, $\beta$. Finally, (\ref{eq:sis_5}) is used to discard spurious solutions.

In the first step, $h_5+h_6$ and $h_3$ are determined from the linear system (\ref{eq:sis_1}), (\ref{eq:sis_4}) as
\begin{equation} \label {eq:sis_6}
 h_5 + h_6   = \frac{\beta} {\alpha - \beta} \, (h_2-k_2) + k_5 + k_6 
\end{equation}
\begin{equation} \label {eq:sis_7}
 h_3    = - \frac{\alpha} {\alpha - \beta} \, (h_2-k_2) + k_3 \ .
\end{equation}
It follows from (\ref{eq:sis_3}) that
\begin{equation} \label {eq:sis_8}
 h_5 \, h_6 = \frac{ k_2}{h_2}  \, k_5 \, k_6 \ .
\end{equation}
The system (\ref{eq:sis_6}),  (\ref{eq:sis_8}) can be solved for the unknowns $h_5$ and $h_6$. The 
resulting pair of solutions  may be expressed as
\begin{equation} \label {eq:sis_9}
(h_5, h_6) = (x_1,x_2) \ ,  \qquad (h_5, h_6) = (x_2,x_1) \ ,
\end{equation}
where 
\begin{equation*} 
x_{1,2}= \frac 12 \, \big[ \frac{\beta} {\alpha - \beta} \, (h_2-k_2) + k_5 + k_6 \pm \sqrt{\Delta}\big] \ ,
\end{equation*}
with
\begin{equation*} 
\Delta =  \big[ \frac{\beta} {\alpha - \beta} \, (h_2-k_2) + k_5 + k_6\big]^2 - 4 \,  \frac{ k_2}{h_2}  \, k_5 \, k_6 \ .
\end{equation*}

In the second step, substitution of equations (\ref{eq:sis_6})--(\ref{eq:sis_8}) into (\ref{eq:sis_2}) provides a third order polynomial equation for $h_2$. After long and tedious calculations, it is written in the form
\begin{equation} \label{eq:pol_h_2}
(h_2-k_2) \, \Big[\frac{\beta^2} {(\alpha - \beta)^2}  \,h_2^2 - \frac {\beta} {\alpha-\beta} \, B \, h_2 +  k_5 \, k_6 \Big] = 0 \ , 
\end{equation} 
where
\begin{equation*}
B=  k_3 - k_5-k_6  + \frac{ \alpha\,} {\alpha-\beta }\, k_2 \ .   
\end{equation*} 
We obtain the three solutions
\begin{equation} \label{eq:sol_h_2}
h_2^{(1)}= k_2 \, \qquad    h_2^{(2)}= \frac {\alpha - \beta}{2 \, \beta} \, (B + \sqrt{\Delta_B}) \ ,  
\qquad    h_2^{(3)}= \frac {\alpha - \beta}{2 \, \beta} \, (B - \sqrt{\Delta_B}) \ ,  
\end{equation} 
with
\begin{equation*}
\Delta_B = B^2 - 4 k_5\,k_6 \ .
\end{equation*} 
In principle, each solution $h_2^{(i)}$, $(i=1, 2, 3)$ generates two vectors $\bh_5$, through substitution into (\ref{eq:sis_7}) and (\ref{eq:sis_9}). 
  
In the third step, we discuss admissibility of the solutions. In general a parameter vector $\bh_5$ can be accepted only if its components are strictly positive. Whenever this condition is not satisfied, the related solution is discarded, and we shall not mention this any more. 
Moreover, any admissible parameter vector $\bh_5$ must satisfy equation (\ref{eq:sis_5}).

Consider the case $h_2=h_2^{(1)}=k_2$. The associated vector parameters $\bh_5^{(1a)}$ and $\bh_5^{(1b)}$ are given by
\begin{equation} \label{eq:bh^1_2}
\bh_5^{(1a)}=(k_1, k_2, k_3, k_5, k_6) \ , \qquad \bh_5^{(1b)}=(k_1, k_2, k_3, k_6, k_5) \ . 
\end{equation}
The vector $\bh_5^{(1a)}$ satisfies equation (\ref{eq:sis_5}), but $\bh_5^{(1b)}$ does not, unless $k_5=k_6$.  
We conclude that $\bh_5^{(1a)}$ is admissible, while $\bh_5^{(1b)}$ is not, if $k_5 \neq k_6$.
Similarly, consider the (positive) components of any vector $\bh_5^{(2a),(2b)}$,  $\bh_5^{(3a),(3b)}$ generated by either $h_2^{(2)}$ or $h_2^{(3)}$; they are expressed in terms of $(k_2,k_3,k_5,k_6)$, $\alpha$, and $\beta$. We say that $\bk_5$ is generic if the corresponding vectors $\bh_5^{(1b)}$, $\bh_5^{(2a),(2b)}$,  $\bh_5^{(3a),(3b)}$ do not satisfy equation (\ref{eq:sis_5}), that is, if they are not admissible. 
Then we can state the following result.

\begin{theorem} \label{teo:id_BCM_tissue}
Assume that the polynomials
\begin{equation*}
Q(s) = \alpha \, s^2 + [\alpha \, ( k_5 + k_6)+ \beta \,k_3]  \, s + \alpha \, k_5 \,k_6  +\beta \, k_3\,k_6 + \gamma \, k_3 \, k_5 \ 
\end{equation*} 
and
\begin{equation*} 
D(s) = s^3 + (k_2+k_3+k_5+k_6) \, s^2 + [ (k_2+k_3) \, (k_5+k_6) + k_5 \, k_6 ] \, s + k_2 \, k_5 \, k_6  \ 
\end{equation*}
are coprime. 
If $\bk_5$ is generic, the rate coefficients $\bk_5 = (k_1, k_2, k_3, k_5, k_6)$ are uniquely determined by $C_i$ and $\mathcal{C}_T$, and the compartmental model of equations (\ref{eq:ode_C}) and (\ref{eq:C_T_tiss_pesato}) is identifiable. 
\end{theorem} 

The proof of identifiability for the BCM dedicated to the kinetics of the cell culture model follows the same lines as the proof of Theorem \ref{teo:id_BCM_tissue}. Therefore, we show here only the main steps. Application of the Laplace transform to equations (\ref{eq:ode_A}) and (\ref{eq:master_eq_act_0}) for the activities, leads to 
\begin{equation} \label{eq:LT_Acells_bis_4C}
\frac{\tilde{\mathcal{A}}_T}{ \tilde{A}_i } =  \frac{\bar{k}_1 \, Q(s)}{D(s)} \ ,
\end{equation}
where
\begin{equation}
Q(s) =  s^2 +  (k_3 + k_5 + k_6) \, s +  (k_3 \, +k_5 )\,k_6 + v \, k_3 \, k_5 \ ,
\end{equation}
and
\begin{equation} 
D(s) = s^3 + (k_2+k_3+k_5+k_6) \, s^2 + [ (k_2+k_3) \, (k_5+k_6) + k_5 \, k_6 ] \, s + k_2 \, k_5 \, k_6  \ .
\end{equation}
Assume that $Q$ and $D$ are coprime. If $\bar{\bh}_5 = (\bar{h}_1,h_2,h_3,h_5,h_6)$ is another vector of rate coefficients consistent with the cell culture data, it follows that $\bar{h}_1=\bar{k}_1$, while the remaining components of $\bar{\bh}_5$ and $\bar{\bk}_5$ satisfy the following system of equations: 
\begin{equation} \label{eq:sis_1_cells}
 h_2+h_3+h_5+h_6=k_2+k_3+k_5+k_6
\end{equation}
\begin{equation} \label{eq:sis_2_cells} 
(h_2+h_3) \, (h_5+h_6) + h_5 \, h_6 = (k_2+k_3) \, (k_5+k_6) + k_5 \, k_6 
\end{equation}
\begin{equation} \label{eq:sis_3_cells}
h_2 \, h_5 \, h_6 = k_2 \, k_5 \, k_6  
\end{equation}
\begin{equation} \label{eq:sis_4_cells}
h_3 + h_5 + h_6=k_3 + k_5 + k_6 
\end{equation}
\begin{equation} \label{eq:sis_5_cells}
(h_3 \, +h_5 )\,h_6 +v \, h_3 \, h_5= (k_3 \, +k_5 )\,k_6 + v\, k_3 \, k_5 \ .
\end{equation}
Comparison between (\ref{eq:sis_1_cells}) and (\ref{eq:sis_4_cells}) shows that $h_2=k_2$. As a consequence, (\ref{eq:sis_3_cells}) reduces to $h_5 \, h_6 = k_5 \, k_6$.  

Next $h_5+h_6$ is determined from (\ref{eq:sis_4_cells}) in terms of $h_3$, and substituted into equation (\ref{eq:sis_2_cells}), which takes the form of a vanishing polynomial of degree 2, in the unknown $h_3$. The corresponding solutions are:
\begin{equation*}
h_3^{(1)} = k_3 \ , \qquad   h_3^{(2)}=  -k_2+k_5+k_6 \ . 
\end{equation*}

If $h_3^{(1)} = k_3$, it is easily shown that $h_5^{(1)} = k_5$ and $h_6^{(1)} = k_6$, which implies $\bar{\bh}_5^{(1)}=\bar{\bk}_5$.

If $h_3^{(2)}\leq 0$ this solution is not admissible. If  $h_3^{(2)}>0$ then equations (\ref{eq:sis_4_cells}) and (\ref{eq:sis_5_cells}) reduce to a linear system for the unknowns $h_5^{(2)}$ and $h_6^{(2)}$. The solution is
\begin{equation*}
h_5^{(2)}= \frac 1 {1-v} \, \big( k_2 - k_3 \, \frac {k_6+v \,k_5}{-k_2+k_5+k_6} \big) \ , \qquad 
h_6^{(2)} = k_2 -  h_5^{(2)} \ . 
\end{equation*}
If at least one between $ h_5^{(2)}$ and $ h_6^{(2)}$ is negative or vanishing, then $h_3^{(2)}$ gives rise to a vector solution which not admissible. If $ h_5^{(2)}$ and $ h_6^{(2)}$ are positive then  the compatibility condition
\begin{equation}  \label{eq:restr}
h_5^{(2)} \, (k_2 -  h_5^{(2)})  =  k_5\, k_6 
\end{equation}
must be satisfied. Thus we conclude that the solution reconstructed from $h_3^{(2)}$ is not admissible, unless the data satisfy equation (\ref{eq:restr}). 

Following the previous procedure, we say that the parameter vector $\bar{\bk}_5$ is generic if it does not satisfy equation (\ref{eq:restr}), and we state the following result.

\begin{theorem} \label{teo:id_BCM_cells}
Assume that the polynomials 
\begin{equation*}
Q(s) =  s^2 +  (k_3 + k_5 + k_6) \, s +  (k_3 \, +k_5 )\,k_6 + v \, k_3 \, k_5 \ 
\end{equation*}
and
\begin{equation*} 
D(s) = s^3 + (k_2+k_3+k_5+k_6) \, s^2 + [ (k_2+k_3) \, (k_5+k_6) + k_5 \, k_6 ] \, s + k_2 \, k_5 \, k_6  \ 
\end{equation*}
are coprime.
If $\bar{\bk}_5$ is generic, the rate coefficients $\bar{\bk}_5 = (\bar{k}_1, k_2, k_3, k_5, k_6)$ are uniquely determined by $A_i$ and $\mathcal{A}_T$, and the compartmental model of equations (\ref{eq:ode_A}) and (\ref{eq:master_eq_act_0}) is identifiable.   
\end{theorem} 

\subsubsection{Estimation of rate constants} \label{subsubsec:GN}

The solution of the compartmental inverse problem for the unknown rate constants requires an optimization-regularization method. Here we describe our approach, based on a Newton-type method, in general terms. For details, see \cite{GN}, \cite{Delbary_GN}, and \cite{Vogel}. This formulation has already been applied successfully in the compartmental framework, e.g. to reduce non-standard compartmental models representing complicated physiologies such as the liver \citep{Garbarino_liver}, to solve the compartmental inverse problem pixelwise in the so-called indirect parametric imaging context \citep{Scusso_IP}, and to address the reference tissue problem of recovering the parameters when the IF is not available \citep{Scusso_RTM}. In these applications the Newton-type method resulted to be rather efficient in the reconstruction of the compartmental kinetic parameters, providing reliable and stable estimates, and performed better than the usual Levenberg-Marquardt method (see Tables 1--3 in \cite{Delbary_GN}, Table III in \cite{Scusso_RTM}).

The underlying ideas of our approach may be described as follows. We rewrite the equation connecting the given data and the compartmental model as a zero finding problem. This means that, for the cell culture, we redefine equation (\ref{eq:master_eq_act_1}) as 
\begin{equation} \label{eq:inv_pb_zero_cells}
\balpha \, \bA(t;\bar{\bk}_5,A_i) - \mathcal{A}_T(t) := \mathcal{F}_t(\bar{\bk}_5) = 0 \ ;
\end{equation}
similarly, equation (\ref{eq:C_T_tiss_pesato_comp}) for the tissue becomes
\begin{equation} \label{eq:inv_pb_zero_tissue}
V_b \, C_i + \balpha \, \bC(t;\bk_5,C_i) - \mathcal{C}_T(t) := \mathcal{F}_t(\bk_5) = 0 \ .
\end{equation}
The vector $\balpha$ is chosen according to the data model.
The input functions $A_i$ and $C_i$ are regarded as given. The total activity of the cell culture $\mathcal{A}_T$ and the total concentration of the target tissue $\mathcal{C}_T$ depend on the unknown vector of parameters $\bar{\bk}_5$ and $\bk_5$, respectively. Notice that equations (\ref{eq:inv_pb_zero_cells}) and (\ref{eq:inv_pb_zero_tissue}) are general enough to hold for both the BCM and SCM, provided that $\bA$ and $\bC$ are substituted with the starred variables $\bA^*$ and $\bC^*$, as in equations (\ref{eq:master_eq_act_*}) and (\ref{eq:C_T_tiss_pesato_Sok}), and the unknown vector of parameters to be considered are $\bar{\bk}_4^*$ and $\bk_4^*$. 

In general, the operator $\mathcal{F}_t: \mathbb{R}^p_+ \to C^1(\mathbb{R}_+, \mathbb{R})$, where $p$ indicates the number of the model coefficients, is a non-linear analytic operator parameterized by the time variable $t \in \mathbb{R}_+$. The Gauss-Newton method transforms the non-linear optimization problem of equation (\ref{eq:inv_pb_zero_cells}), or equation (\ref{eq:inv_pb_zero_tissue}), into a linear problem by computing the Fr\`echet derivative of the operator $ \mathcal{F}_t$ with respect to the kinetic parameters. What is found is a linear equation 
\begin{equation} \label{eq:GN}
\bigg[ \frac{d \mathcal{F}_t}{d \bk} (\bk^{(0)};\bh^{(0)}) \bigg](t) = - \mathcal{F}_t(\bk^{(0)}) \ ,
\end{equation}
with the bounded and linear differential operator $d \mathcal{F}_t/d \bk$, unknown step-size $\bh^{(0)} \in \mathbb{R}^p$, initial guess $\bk^{(0)} \in \mathbb{R}_+^p$, and for $t \in \mathbb{R}_+$.
In real applications, only noisy versions of the data for a finite number of sampling time points $t_1, \dots, t_n \, \in \mathbb{R}_+$ are available. Therefore, equation (\ref{eq:GN}) becomes the discretized linear system
\begin{equation} \label{eq:GN_discr}
\boldsymbol{F}_0 \, \bh^{(0)} = \boldsymbol{Y}_0 \ ,
\end{equation} 
where $\boldsymbol{F}_0$ is the matrix encoding the Fre\`chet derivatives with respect to $\bk^{(0)}$, and $\boldsymbol{Y}_0$ is the vector discretizing $- \mathcal{F}_t$ computed in $\bk^{(0)}$.
The system (\ref{eq:GN_discr}) constitutes a classic linear ill-posed inverse problem, since the solution may not exist, may not be unique, and may not be stable. In order to find a unique stable solution of (\ref{eq:GN_discr}), we consider a Tikhonov-type regularization, with the Tikhonov penalty on the step-size vector, which leads to the regularized system
\begin{equation} \label{eq:GN_discr_reg}
( \boldsymbol{F}_0^T \, \boldsymbol{F}_0 + \lambda_0 \, \boldsymbol{I}_{[p]} ) \, \bh^{(0)} = \boldsymbol{F}_0^T \, \boldsymbol{Y}_0 \ ,
\end{equation} 
where $\boldsymbol{I}_{[p]}$ is the identity matrix of dimension $p$, and $\lambda_0$ is the regularization parameter which is allowed to change at every iteration. The regularization parameter may be fixed \emph{a priori}, or selected with a proper method, e.g. the Generalized Cross Validation (GCV) method \citep{GCV}. 
The optimization algorithm performs an iterative scheme which: 1) starts from a random initial guess $\bk^{(0)}$, 2) determines the step-size $\bh^{(0)}$ as the least-square solution of (\ref{eq:GN_discr_reg}), 3) updates the values of the kinetic parameters by letting $\bk^{(1)} = \bk^{(0)} + \bh^{(0)}$ and 4) iterates the process.
To stop the iterative algorithm, we check the relative error between the given experimental datum and the model-predicted one, using a threshold coinciding with the uncertainty on the measurement as a stopping criterion. 

\section{Applications to cancer cell cultures \emph{in vitro}} \label{sec:LT}  

In this section we determine the rate coefficients describing FDG kinetics of cultures of 4T1 cancer cells (breast cancer cell lines), and the corresponding compartment activities. The data have been obtained by the use of a LigandTracer (LT) device of Ridgeview Instruments \citep{Bjorke1,Bjorke2,Mertens}. Details on experimental procedures and calibration methods applied in order to follow tracer uptake by cell cultures can be found in the forthcoming paper (Scussolini et al. manuscript in preparation). Here, we make use of new experimental data.

Application of the BCM to cell cultures is in natural relation with the cell origin of the model. Cell cultures allow for repeated experiments under constant conditions, whereas experiments on tracer uptake \emph{in vivo} may be influenced by absorption by other organs, specific tissue environment, blood perfusion, and so on. 
Moreover, LT-measurements allow a direct estimate of FDG consumption, and thus of glucose consumption, without any distortion introduced by physical corrections or signal reconstruction algorithm, which are essential steps to be made in, e.g., PET experiments \emph{in vivo}. For this reasons, the data coming from LT-cells experiments are highly stable and reliable with respect to cell biology, and the results give a fair interpretation of the phenomenon observed.

Comparison with the results obtained from the analysis of cancer tissues, described in the next section, provides a deep understanding of the feasibility and effectiveness of the compartmental model. Contrast with results available in the literature is obtained through application of the SCM to data reduction. 

\subsection{Data} \label{subsec:LT_dev}   

\begin{figure}[htb]
\centering
\subfigure[LT device. \label{fig:LT}]
{\includegraphics[width=6.5cm]{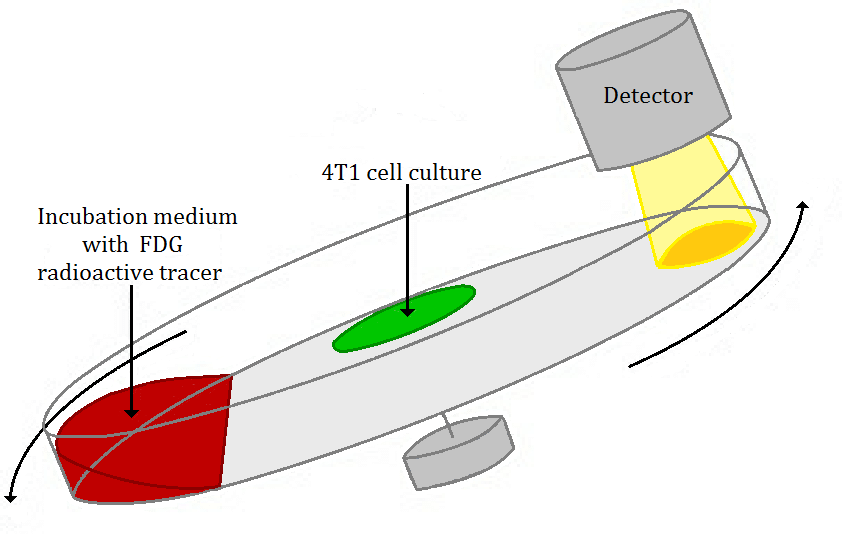}} \\ 
\subfigure[$\mathcal{A}_T$. \label{fig:AT}]
{\includegraphics[width=5.5cm]{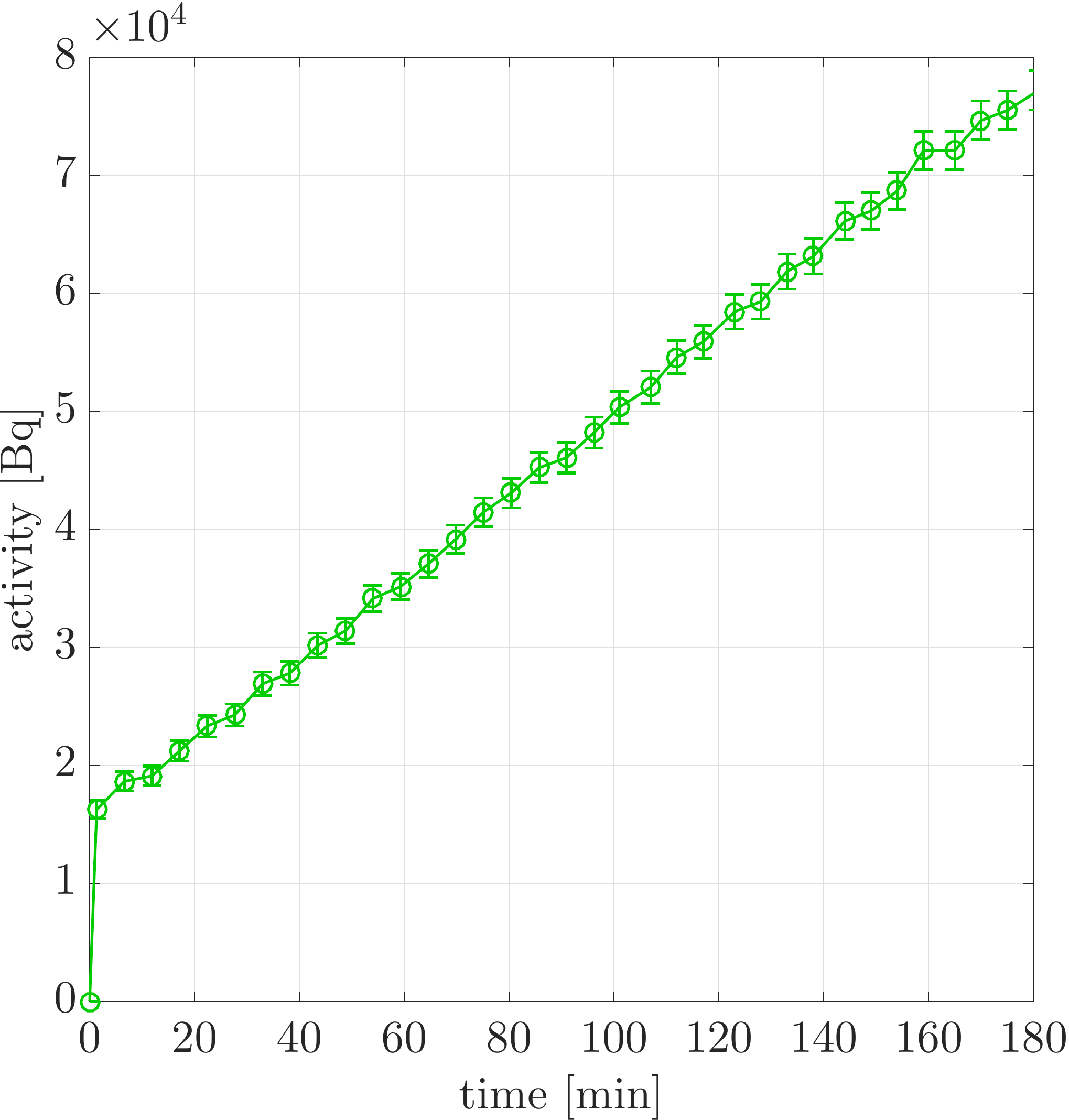}} \
\subfigure[$A_i$. \label{fig:Ai}]
{\includegraphics[width=5.8cm]{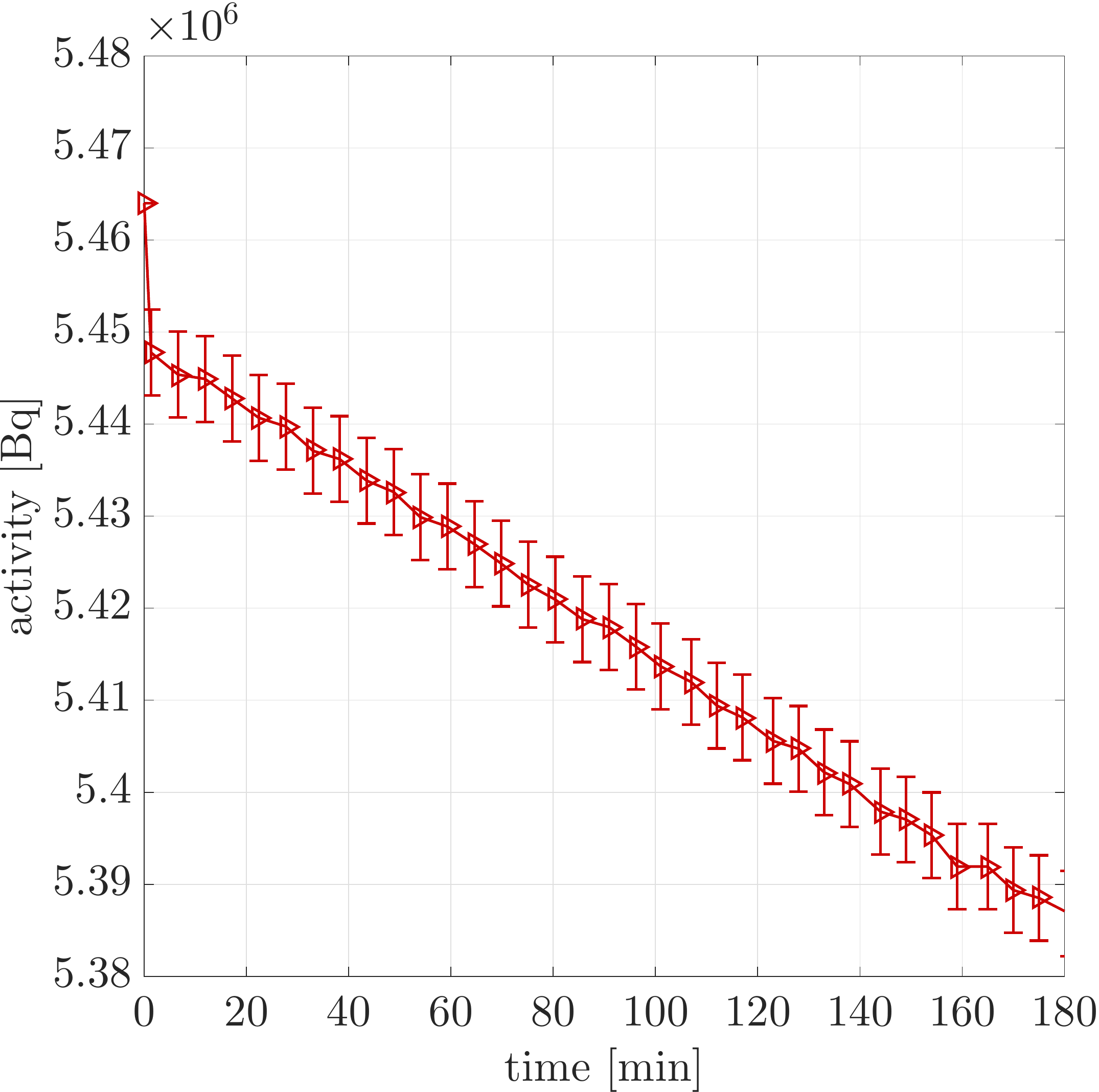}} 
\caption{(a) Measurement principle of the LT device: the petri dish containing attached target cells is placed on an inclined and rotating support; the incubation medium with the FDG radioactive tracer occupies the lower part of the dish due to the dish inclination; the detector points towards the upper part of the dish and the uptake of FDG by cells is measured once per rotation in the upper position. (b) The time-dependent activity curve of FDG uptake $\mathcal{A}_T$ and its standard deviation, related to experiment e1. (c) The time-dependent activity curve of the incubation medium $A_i$ and its standard deviation, related to experiment e1} 
\label{fig:LT_data}
\end{figure}

Cultured 4T1 cancer cells have been seeded and then attached over a specific portion of the surface of a petri dish held by the LT device (see \figurename~\ref{fig:LT}).  
The bottom of the dish has been filled with a radioactive incubation medium containing both glucose at physiological concentration 1 g/L, i.e. 5.5mM, and an amount of FDG corresponding to about $10^6$ Bq, diluted in a volume of 3 mL. Notice that the amount of FDG can be considered negligible with respect to that of glucose, i.e., FDG has to be regarded as a perturbation of glucose.
The dish has been subject to a periodic motion around its axis, inclined from the vertical; at each rotation cycle, lasting one minute, the LT device has collected the radioactivity emitted by the cells. Experiments have been performed for a total time interval of 180 minutes. 

In the course of a typical experiment, radioactive FDG molecules, initially added to the incubation medium, have been uptaken and then retained by the cell culture. 
For each experiment we have considered the time dependent total activity (Bq) of the cell culture $\mathcal{A}_T$, and the corresponding input function $A_i$, 
describing the activity inside the incubation medium. Both activity curves have been decay corrected.  
Since the LT is a closed system for radioactive molecules, the two curves satisfy the conservation law  $A_i + \mathcal{A}_T= A_{i0}$, where the known constant $A_{i0}$ represents the activity in the medium available at the beginning of the measuring procedure, after absorption by wet surfaces, which occurs in a very short time interval.  
The total activity $\mathcal{A}_T$ has been reconstructed on the basis of the counts of the detector available with the LT. The input function has been determined by the conservation law as $A_i = A_{i0}-\mathcal{A}_T$.        
 
Following \cite{Milo}, we have chosen the value of the intracellular relative size of the ER with respect to the cytosol as $v = 0.17$, which holds for a rough ER in a liver hepatocyte cell.

We have considered six LT experiments, denoted as e$i$, with $i =1,\dots,6$, differing between each other for number of cells $N_c$ and initial amount of FDG in the medium $A_{i0}$. 
\tablename~\ref{tab:exp_data_cells} reports the experimental values of the number of cells $N_c$, the initial amount of FDG in the medium $A_{i0}$, the end-time total activity of the cell culture $\mathcal{A}_T$, and the slope (Bq/min) of the line approximating $\mathcal{A}_T$ (by means of linear regression, with a coefficient of determination $r^2$ oscillating between $0.97$ and $0.99$), as an estimate of the growth rate of the activity of cells.

\begin{table}
\caption{Experimental values of the number of cells $N_c$, the initial FDG activity in the medium $A_{i0}$ (Bq), the final total activity of cells $\mathcal{A}_T(180)$ (Bq), and the growth rate of $\mathcal{A}_T$ as the slope (Bq/min) of the line approximating the curve, for each LT experiment. Notice that [M] refers to multiplication by $10^6$}
\label{tab:exp_data_cells}
\begin{tabular}{lllll}
\hline\noalign{\smallskip}
& $N_c$ [M] & $A_{i0}$ [M] & $\mathcal{A}_T(180)$ & growth rate \\
\noalign{\smallskip}\hline\noalign{\smallskip}
e1 & $0.96$ & $5.46$ & $7.72 \cdot 10^4$ & $348$ \\ 
e2 & $0.40$ & $5.26$ & $7.85 \cdot 10^4$ & $434$ \\ 
e3 & $0.40$ & $5.46$ & $6.61 \cdot 10^4$ & $305$ \\ 
e4 & $0.80$ & $6.39$ & $1.01 \cdot 10^5$ & $545$ \\ 
e5 & $0.80$ & $8.37$ & $9.16 \cdot 10^4$ & $412$ \\ 
e6 & $0.60$ & $4.74$ & $2.26 \cdot 10^4$ & $82$ \\ 
\noalign{\smallskip}\hline
\end{tabular}
\end{table}
 
As typical example of time-dependent activity curve of FDG uptake by the cell culture, \figurename~\ref{fig:AT} shows the datum $\mathcal{A}_T$ of experiment e1.
In general, the graph of $\mathcal{A}_T$ exhibits a certain degree of variability among the experiments because of the the different experimental setup. Nevertheless, the qualitative behavior of the uptake curves is relatively well defined: at each experiment $\mathcal{A}_T$ grows almost linearly, with small random oscillations that should be due to experimental errors. A similar behavior had already been observed both \emph{in vitro} and \emph{in vivo} (see, e.g., \cite{Mertens} and references cited therein).

An example of input function can be seen in \figurename~\ref{fig:Ai} (experiment e1). The graph of $A_i$ is almost constant, in that the relative loss of tracer from the medium with respect to the initial amount, in the total time-interval of 180 min, is about 1$\%$; in other terms, the cell culture uptake of tracer from the incubation medium is small with respect to the total amount of tracer in the medium.

\subsection{Results} \label{subsec:LT_results}

We have analyzed 4T1 cell culture data with both the BCM and SCM. The results are reported in \tablename~\ref{tab:k_cells_BCM} for the BCM reduction, and in \tablename~\ref{tab:k_cells_SCM} for the SCM reduction. 
Means and standard deviations have been computed over 50 runs of the iterative algorithm, with different initialization of the kinetic parameters, randomly chosen in the interval $(0,1)$ with uniform distribution.
The regularized Gauss-Newton algorithm is rather robust with respect to the choice of the regularization parameter, as showed in \cite{Delbary_GN}; in this application the regularization parameter has been  fixed for each iteration at the value of $10^6$.
The iterative algorithm was stopped when the relative error between the experimental activity and the model-predicted one, computed with the L$^2$ norm, was lower than a threshold of the order of $10^{-2}$. 

\begin{table}
\caption{Reconstructed kinetic parameters (1/min) by the use of the BCM for the 4T1 cell culture of the LT experimental group of six experiments, as mean and standard deviation over 50 runs of the Gauss-Newton algorithm. The last two lines report mean and standard deviation of each kinetic parameter computed over the mean estimates of the six experiments}
\label{tab:k_cells_BCM}
\resizebox{\textwidth}{!}{
\begin{tabular}{llllll}
\hline\noalign{\smallskip}
& $\bar{k}_1$ & $k_2$ & $k_3$ & $k_5$ & $k_6$ \\
\noalign{\smallskip}\hline\noalign{\smallskip}
e1 & $0.0083 \pm 0.0007$ & $2.8722 \pm 0.2710$ & $0.1340 \pm 0.0021$ & $2.0803 \pm 0.3009$ & $0.0000 \pm 0.0000$ \\ 
e2 & $0.0073 \pm 0.0012$ & $4.0378 \pm 0.7763$ & $0.3396 \pm 0.0022$ & $0.7936 \pm 0.0099$ & $0.0021 \pm 0.0000$ \\ 
e3 & $0.0050 \pm 0.0006$ & $2.4056 \pm 0.3527$ & $0.1939 \pm 0.0030$ & $2.2454 \pm 0.6789$ & $0.0014 \pm 0.0000$ \\ 
e4 & $0.0152 \pm 0.0018$ & $6.3428 \pm 0.8059$ & $0.2497 \pm 0.0434$ & $3.7148 \pm 0.8222$ & $0.0354 \pm 0.1702$ \\ 
e5 & $0.0153 \pm 0.0012$ & $7.1280 \pm 0.5583$ & $0.1510 \pm 0.0002$ & $4.5027 \pm 0.1900$ & $0.0009 \pm 0.0000$ \\ 
e6 & $0.0067 \pm 0.0001$ & $5.1039 \pm 0.1030$ & $0.0948 \pm 0.0002$ & $2.7223 \pm 0.2495$ & $0.0020 \pm 0.0000$ \\ \\
mean & $0.0096$ & $4.6484$ & $0.1938$ & $2.6765$ & $0.0070$ \\ 
std & $0.0045$ & $1.8860$ & $0.0890$ & $1.3040$ & $0.0140$ \\ 
\noalign{\smallskip}\hline
\end{tabular}}
\end{table}

\begin{table}
\caption{Reconstructed kinetic parameters (1/min) by the use of the SCM for the 4T1 cell culture of the LT experimental group of six experiments, as mean and standard deviation over 50 runs of the Gauss-Newton algorithm. The last two lines report mean and standard deviation of each kinetic parameter computed over the mean estimates of the six experiments}
\label{tab:k_cells_SCM}
\begin{tabular}{lllll}
\hline\noalign{\smallskip}
& $\bar{k}_1^*$ & $k_2^*$ & $k_3^*$ & $k_4^*$ \\
\noalign{\smallskip}\hline\noalign{\smallskip}
e1 & $0.0078 \pm 0.0011$ & $2.6919 \pm 0.3771$ & $0.0222 \pm 0.0006$ & $0.0000 \pm 0.0000$ \\ 
e2 & $0.0041 \pm 0.0017$ & $1.7424 \pm 0.7908$ & $0.0426 \pm 0.0010$ & $0.0020 \pm 0.0000$ \\ 
e3 & $0.0048 \pm 0.0008$ & $2.3039 \pm 0.3913$ & $0.0307 \pm 0.0001$ & $0.0013 \pm 0.0000$ \\ 
e4 & $0.0144 \pm 0.0001$ & $5.8915 \pm 0.0450$ & $0.0417 \pm 0.0000$ & $0.0020 \pm 0.0000$ \\ 
e5 & $0.0150 \pm 0.0002$ & $6.9155 \pm 0.0966$ & $0.0250 \pm 0.0000$ & $0.0010 \pm 0.0000$ \\ 
e6 & $0.0068 \pm 0.0000$ & $5.1015 \pm 0.0318$ & $0.0157 \pm 0.0000$ & $0.0019 \pm 0.0000$ \\ \\
mean & $0.0088$ & $4.1078$ & $0.0296$ & $0.0014$ \\ 
std & $0.0048$ & $2.1404$ & $0.0108$ & $0.0008$ \\ 
\noalign{\smallskip}\hline
\end{tabular}
\end{table} 

The following comments to the results of \tablename~\ref{tab:k_cells_BCM} and \tablename~\ref{tab:k_cells_SCM} are in order.
\begin{itemize}

\item At each experiment, the reconstructed values of $\bar{k}_1$ and $\bar{k}_1^*$ show only slight numeric differences and are very small (order of magniture $10^{-2}$). We recall that $\bar{k}_1$ was defined as $\bar{k}_1= k_1 \, \mathscr{V}_\text{cyt}/ \mathscr{V}_\text{i}$, with $\mathscr{V}_\text{cyt} \ll  \mathscr{V}_\text{i}$, which implies that the smallness of $\bar{k}_1$ is ultimately related to the choice of activities as state variables. Of course, the contribution $\bar{k}_1 \,A_i$ cannot be discarded from the system (\ref{eq:ode_A}) because it is of the order of $10^4$. Similar remarks apply to $\bar{k}_1^*$. 

\item The estimated values of $\bar{k}_2$ and $\bar{k}_2^*$ are almost equal and of order of unity. 

\item Taking into account also the activity curves, it may be shown that the overall contribution $-k_2 \,A_f + \bar{k}_1 \,A_i$ to the time rate $\dot{A}_f$, due to FDG exchange between incubation medium and cytosol, is strictly positive (as expected) but rather small. This is consistent with the small decrease rate in time of the activity $A_i$ of the incubation medium, and the expectation that only a small fraction of the FDG contained in the medium is consumed by the system of cells.    

\item The result that $\bar{k}_1 \approx \bar{k}_1^*$ and $k_2 \approx k_2^*$ shows that the two rate constants cannot be used to discriminate between the two models BCM and SCM. This also implies that the reconstructed tracer exchange between cells and incubation medium is independent of the model applied. 

\item The estimated values of $k_6$ result of the order of $10^{-3}$ and are almost coincident with those of the corresponding parameter $k_4^*$ of the SCM. They can be set equal to 0, as it is often done, following \cite{Sokoloff}.

\item The estimated values of $k_3$ are greater than those of $k_3^*$, implying a different value for the phosphorylation rate predicted by the competing models. We also observe that the assumptions made in subsection \ref{subsec:BCM_vs_SCM} may be considered as  satisfied by the reconstructed parameters and related compartment activities; indeed, the reconstructed values satisfy the relation $k_3^* \approx v k_3$, with $v=0.17$, which is a particular case of (\ref{eq:id_k_cells}). From a different viewpoint, this shows the reliability of the inversion procedure.

\end{itemize}

\begin{figure}[htb]
\centering
\subfigure[BCM. \label{fig:BCM_comp_cells}]
{\includegraphics[width=5.5cm]{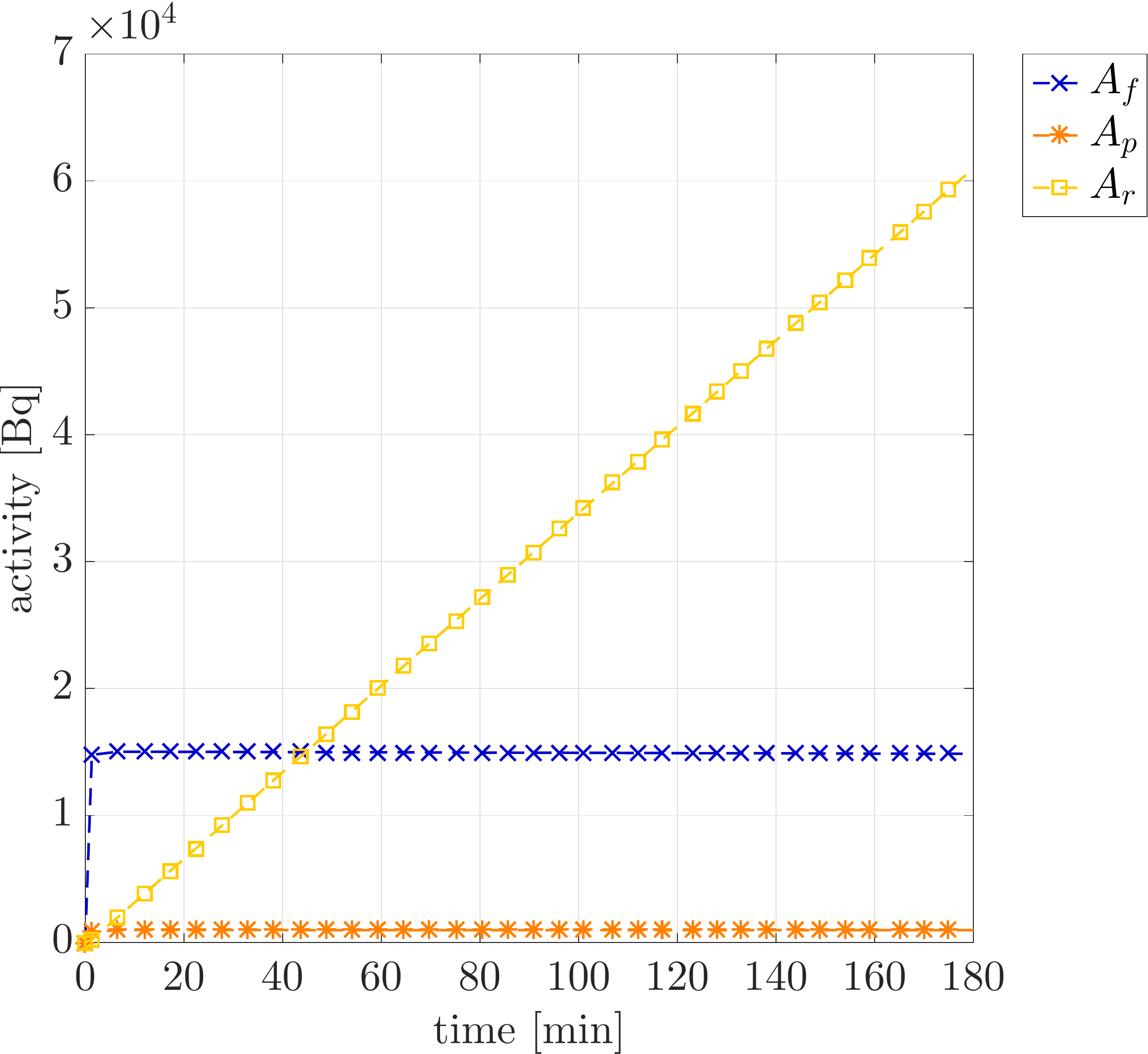}} \
\subfigure[SCM. \label{fig:SCM_comp_cells}]
{\includegraphics[width=5.5cm]{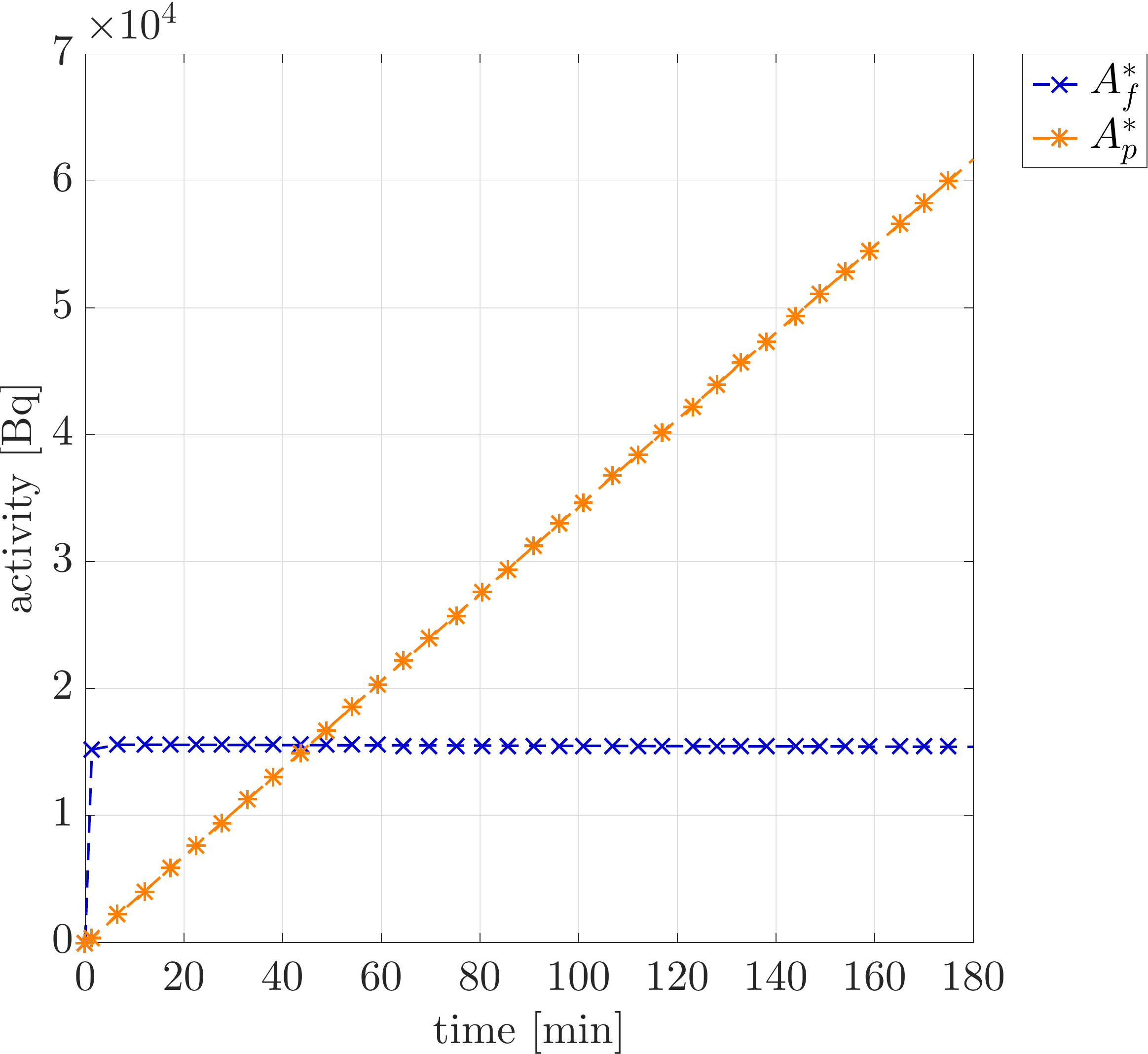}} 
\caption{Model-predicted time curves of the compartment activities for the experiment e1: (a) $A_f$, $A_p$ and $A_r$ of the BCM; (b) $A_f^*$, and $A_p^*$ of the SCM} 
\label{fig:comp_cells}
\end{figure}

\figurename~\ref{fig:BCM_comp_cells} shows the reconstructed time-activity curves of the BCM compartments for the experiment e1, as representative of all experiments conducted. It is immediately evident that the FDG is accumulated in the ER compartment; in fact, the ER activity $A_r$ increases in time almost linearly and reaches the maximum value at the end-time point. 
The free tracer activity $A_f$ is almost constant, with stationary value reached in the first few minutes of the experiment.
The cytosolic phosphorylated tracer $A_p$ is approximately constant, and it is almost one order of magnitude smaller than $A_f$, showing that a small (constant) amount of phosphorylated FDG occupies the cytosol, where the amount of free tracer prevails over that of phosphorylated; by the way, this also indicates a high efficiency of the process of transfer of tracer molecules to ER.
For comparison, in \figurename~\ref{fig:SCM_comp_cells} the time-activity curves of the reconstructed SCM compartments for experiment e1 are shown. Again, the compartment $A_f^*$ for free tracer becomes asymptotically stable in the first minutes, while the compartment $A_p^*$ for phosphorylated tracer contains the greater amount of radioactive molecules and represents the pool where the FDG is accumulated. 

\section{Applications to cancer  tissues \emph{in vivo}} \label{sec:PET}

In this section, the BCM and the SCM approaches are applied to the reduction of the same cancer tissue data, in order to give evidence to differences in the reconstructed kinetics. In the first step, we compare results obtained by the action of the two alternative compartmental models on the same set of synthetic data. 
In the second step, we consider application of the two models to real FDG--PET data of cancer murine models. This allows in particular comparison with reconstructions of FDG kinetics in tumor tissue available in the literature, which have been obtained by application of the Sokoloff-type two-compartment model.

\subsection{Validation on simulation setting} \label{subsec:PET_simulation}
In order to test the proposed BCM against the standard SCM, we have generated tissue data by a standard procedure: choice of a realistic IF, selection of realistic ground-truth values for the rate constants, solution of the related direct problem for BCM, reconstruction of the total concentration. Then, the inverse problem has been solved by application of the BCM and SCM, with the specific aim of analyzing the change in the numerical values of the reconstructed parameters induced by change of the model applied for reduction. 

Synthetic data have been produced by using 27 time frames equivalent to the typical total acquisition time of the FDG experiments performed with the microPET scanner ``Albira'' available at our lab (Carestream Health, Genova, user manual by \cite{Albira}), and in agreement with usual time points of the experiments ($10 \times15$ sec + $1 \times 22$ sec + $4 \times30$ sec + $5 \times 60$ sec + $2 \times 150$ sec + $5 \times 300$ sec). The arterial IF has been simulated by fitting with a gamma variate function \citep{Golish} a set of real measurements acquired from a healthy mouse in a controlled experiment. With the ground-truth values of the BCM parameters $\bk_5 = (k_1, k_2, k_3, k_5, k_6)$, and with the synthetic IF, the state variables have been evaluated by means of equation (\ref{eq:sol_C}). The total tissue concentration $\mathcal{C}_T$ has been computed by equation (\ref{eq:C_T_tiss_pesato_comp}), where the values of the volume fractions have been fixed as $V_b = 0.15$, $V_i = 0.3$, and $v=0.17$ so that $v_r = v/(1+v) = 0.14$. 

We have created fifty independent identically-distributed noisy datasets by adding to $\mathcal{C}_T$ white Gaussian noise with a signal-to-noise ratio of 30 dB, producing realistic signals for the activity of radio-tracer in tissues. For each dataset, we have solved the inverse problem of equation (\ref{eq:inv_pb_zero_tissue}) for the unknown vector of parameters $\bk_5$, by applying the BCM, and for the unknown vector $\bk_4^*$, by applying the SCM. In so doing we have performed a model-sensitivity analysis, as well as we have tested the Gauss-Newton algorithm reliability for BCM.
The starting point of the iterative method has been randomly chosen in the interval (0,1), and the regularization parameter has been optimized at each iteration through the GCV method \citep{GCV}, by the requirement of a predefined range of variability (between $10^2$ and $10^4$). The algorithm has been stopped when the relative error between the original noisy total concentration and the model-predicted one, computed with the L$^2$ norm, has become lower than a threshold of order of $10^{-2}$.  
The ground-truth values and the reconstructed values of the parameters for both the BCM and the SCM are reported in \tablename~\ref{tab:k_simul_BCM_SCM}. Means and standard deviations are computed over the fifty different realizations. Notice that in \tablename~\ref{tab:k_simul_BCM_SCM} we do not indicate explicitly the notation $k^*$ for the SCM coefficients, but, with a slight abuse of language, we identify the BCM and SCM parameters with the same kinetic meaning, i.e. $k_1$ with $k_1^*$, $k_2$ with $k_2^*$, $k_3$ with $k_3^*$, and $k_6$ with $k_4^*$. 

The following comments to \tablename~\ref{tab:k_simul_BCM_SCM} are now in order.
\begin{itemize}
\item The procedure for the solution of the inverse problem may be considered as sufficiently reliable: in particular, the reduction of the BCM by means of the Gauss-Newton algorithm provides accurate reconstructions of the ground-truth values with rather small standard deviations.
\item Comparison of $k_1^*$ and $k_4^*$ with $k_1$ and $k_6$, respectively, shows that change of the model in the reduction procedure has a negligible influence on these reconstructed values.
\item The $k_2^*$ value returned by the SCM overestimates the $k_2$ ground-truth value, while the $k_3^*$ value underestimates the $k_3$ ground-truth value.  
\end{itemize}
To summarize, differences in the reconstructed parameter values resulting from application of the BCM and the SCM are mainly concerned with the pairs $(k_2,k_2^*)$ and $(k_3,k_3^*)$; the gaps originate from the non equivalent choices of the number of state variables in the two models, and the different expressions of the total concentration $\mathcal{C}_T$, either given by (\ref{eq:C_T_tiss_pesato_comp}) or (\ref{eq:C_T_tiss_pesato_Sok}).

\begin{table}
\caption{Ground-truth and reconstructed values for the kinetic parameters (1/min) of the BCM and of the SCM by the use of the Gauss-Newton method, as mean and standard deviation over 50 different runs of the algorithm. The parameters of the BCM and of the SCM with the same kinetic meaning are identified: $k_1$ with $k_1^*$, $k_2$ with $k_2^*$, $k_3$ with $k_3^*$, and $k_6$ with $k_4^*$}
\label{tab:k_simul_BCM_SCM}
\begin{tabular}{llllll}
\hline\noalign{\smallskip}
& $k_1$ & $k_2$ & $k_3$ & $k_5$ & $k_6$ \\
\noalign{\smallskip}\hline\noalign{\smallskip}
ground-truth & $0.4$ & $0.2$ & $0.7$ & $0.5$ & $0.03$ \\
BCM & $0.40 \pm 0.02$ & $0.22 \pm 0.07$ & $0.69 \pm 0.18$ & $0.49 \pm 0.14$ & $0.03 \pm 0.03$ \\
SCM & $0.36 \pm 0.01$ & $0.44 \pm 0.03$ & $0.11 \pm 0.05$ & $-$ & $0.02 \pm 0.03$ \\  
\noalign{\smallskip}\hline
\end{tabular}
\end{table}    

\subsection{Analysis of real data \emph{in vivo}} \label{subsec:PET_realdata}

\begin{figure}[htb]
\centering
\subfigure[ROIs. \label{fig:ROIs}]
{\includegraphics[width=5.5cm]{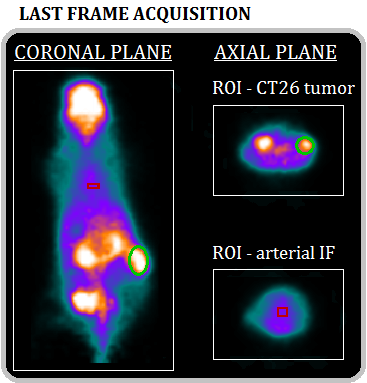}} \\
\subfigure[$\mathcal{C}_T$. \label{fig:CT}]
{\includegraphics[width=5.5cm]{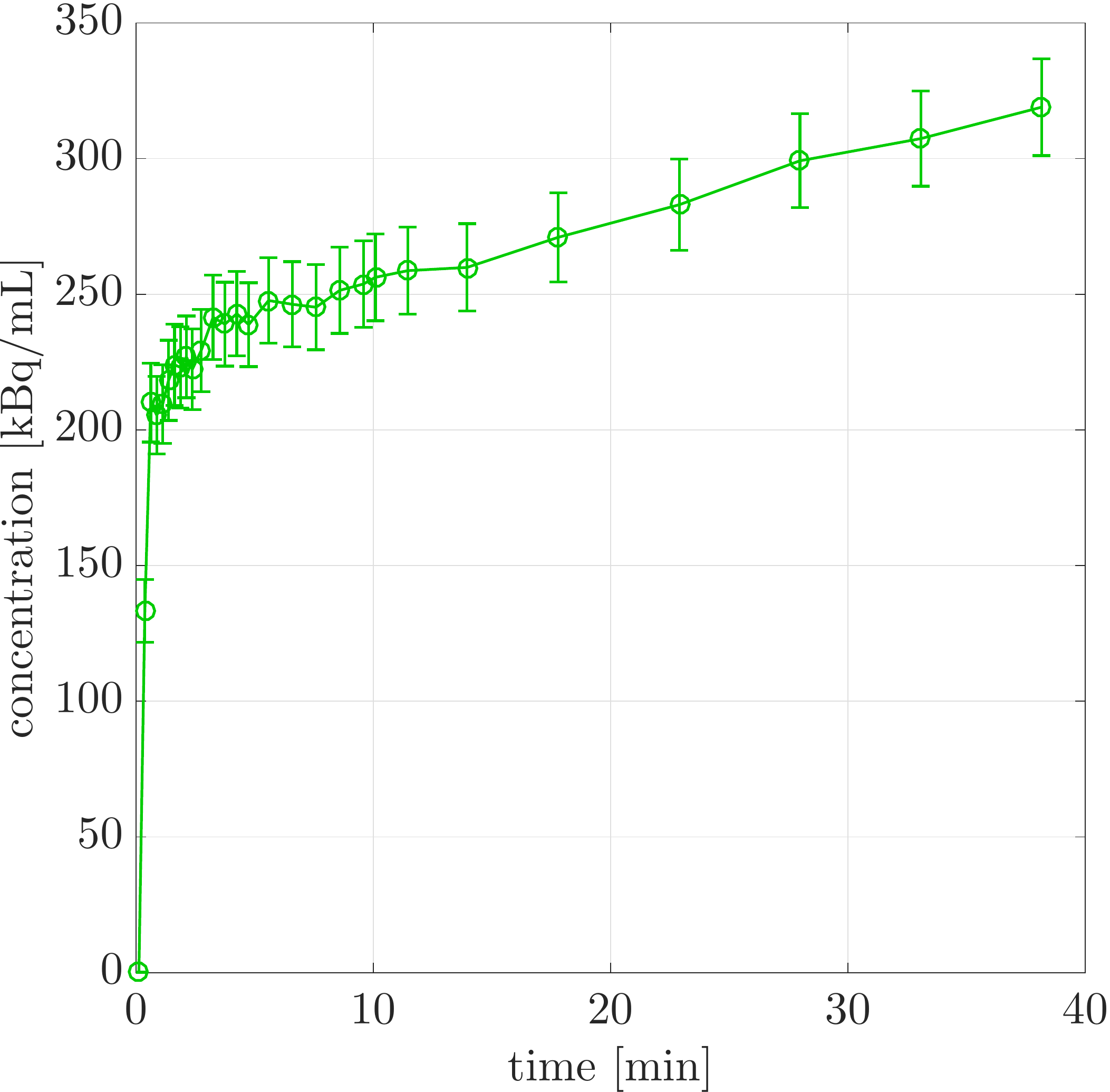}} \
\subfigure[$C_i$. \label{fig:Ci}]
{\includegraphics[width=5.6cm]{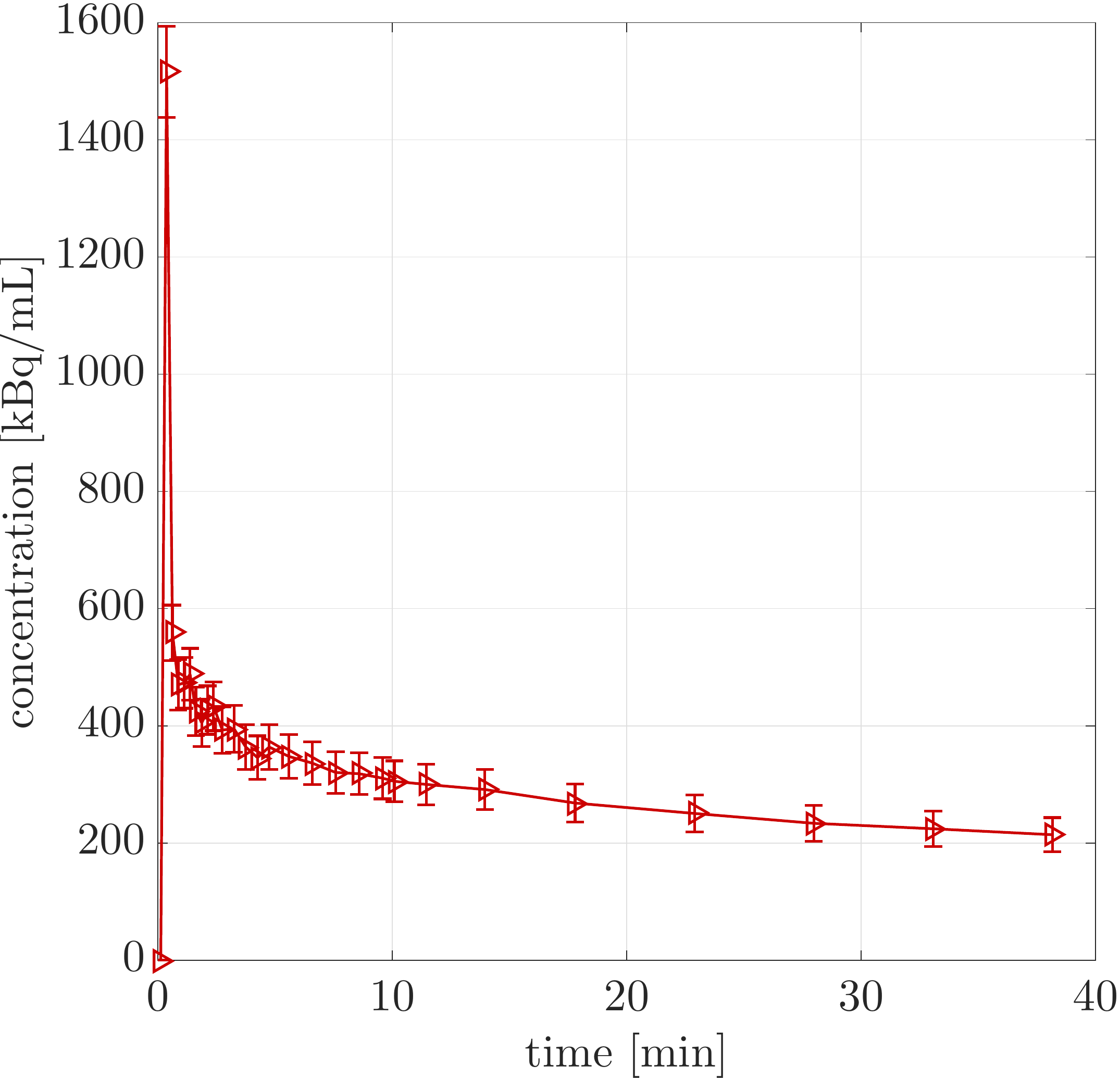}} 
\caption{(a) Last frame of the FDG--PET acquisition of the murine model m1 with ROIs around the CT26 tumor (green color) and the aortic arc (red color). (b) The time-dependent ROI concentration curve of the CT26 tumor $\mathcal{C}_T$  and its standard deviation, related to experiment m1. (c) The time-dependent concentration curve of the arterial input function $C_i$  and its standard deviation, related to experiment m1} 
\label{fig:tissue_data}
\end{figure}

FDG--PET real data of murine models have been obtained by means of a dedicated microPET system (Albira, Carestream Health, Genova, described in \cite{Albira}), currently operational at our lab (IRCCS San Martino IST, Genova). The experimental protocol for FDG--PET experiments has followed the steps described in \cite{protocol}. In particular, all animals have been studied after a fasting period of six hours, to ensure a steady state of substrate and hormones governing glucose metabolism, and have been properly anesthetized and positioned on the bed of the microPET system, whose two-ring configuration covers the whole animal body in a single bed position. A dose of 3 to 4 MBq of FDG has been injected through a tail vein, soon after the start of a dynamic list mode acquisition lasting 40 min. The acquisition has been reconstructed using the framing rate $10 \times15$ sec + $1 \times 22$ sec + $4 \times30$ sec + $5 \times 60$ sec + $2 \times 150$ sec + $5 \times 300$ sec, and then PET data have been reconstructed using a Maximum Likelihood Expectation Maximization (MLEM) method \citep{Shepp}.
Animals have been inoculated subcutaneously in the dorsal hip muscles with $2 \cdot 10^5$ murine cancer cell lines CT26 (colon carcinoma cell lines).

To obtain the Region Of Interest (ROI) concentrations (kBq/mL), each image dataset has been reviewed by an experienced observer who recognized two ROIs: one over the cancer lesion to compute $\mathcal{C}_T$, and one over the left ventricle in order to compute the blood IF $C_i$. 
The blood volume fraction has been set equal to $V_b = 0.15$, according to \cite{Montet} for tumor in CT26-tumor bearing mice, and the interstitial volume fraction has been chosen as $V_i = 0.3$, following \cite{Kim}. The relative size of the ER with respect to the cytosol has been imposed as $v=0.17$, as for the cell cultures. Therefore, the volume fraction of the ER with respect to the cell environment has been fixed as $v_r = 0.14$.

\begin{figure}[htb]
\centering
\subfigure[BCM. \label{fig:K_BCM_tissue}]
{\includegraphics[width=5cm]{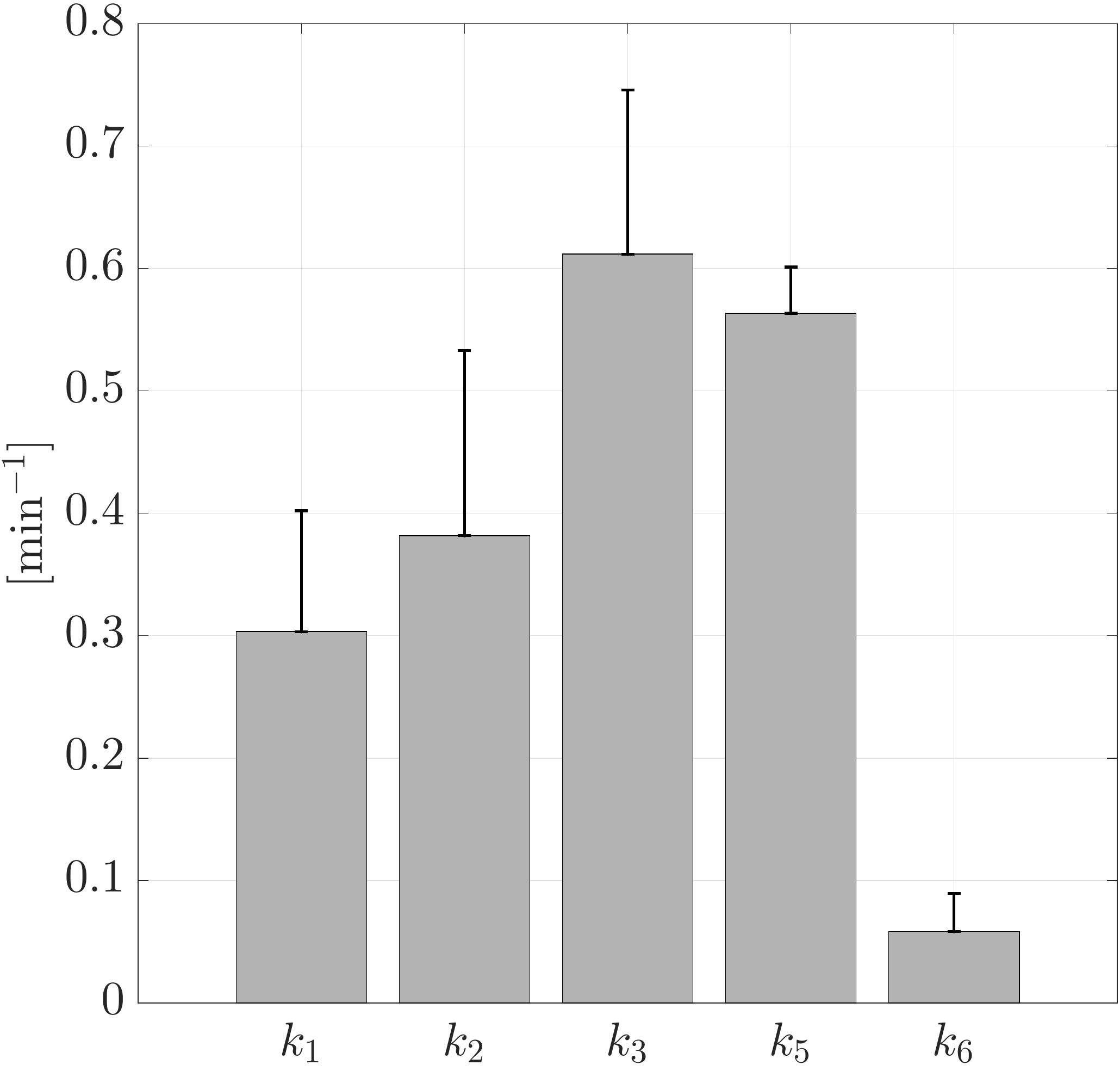}} \
\subfigure[SCM. \label{fig:K_SCM_tissue}]
{\includegraphics[width=5cm]{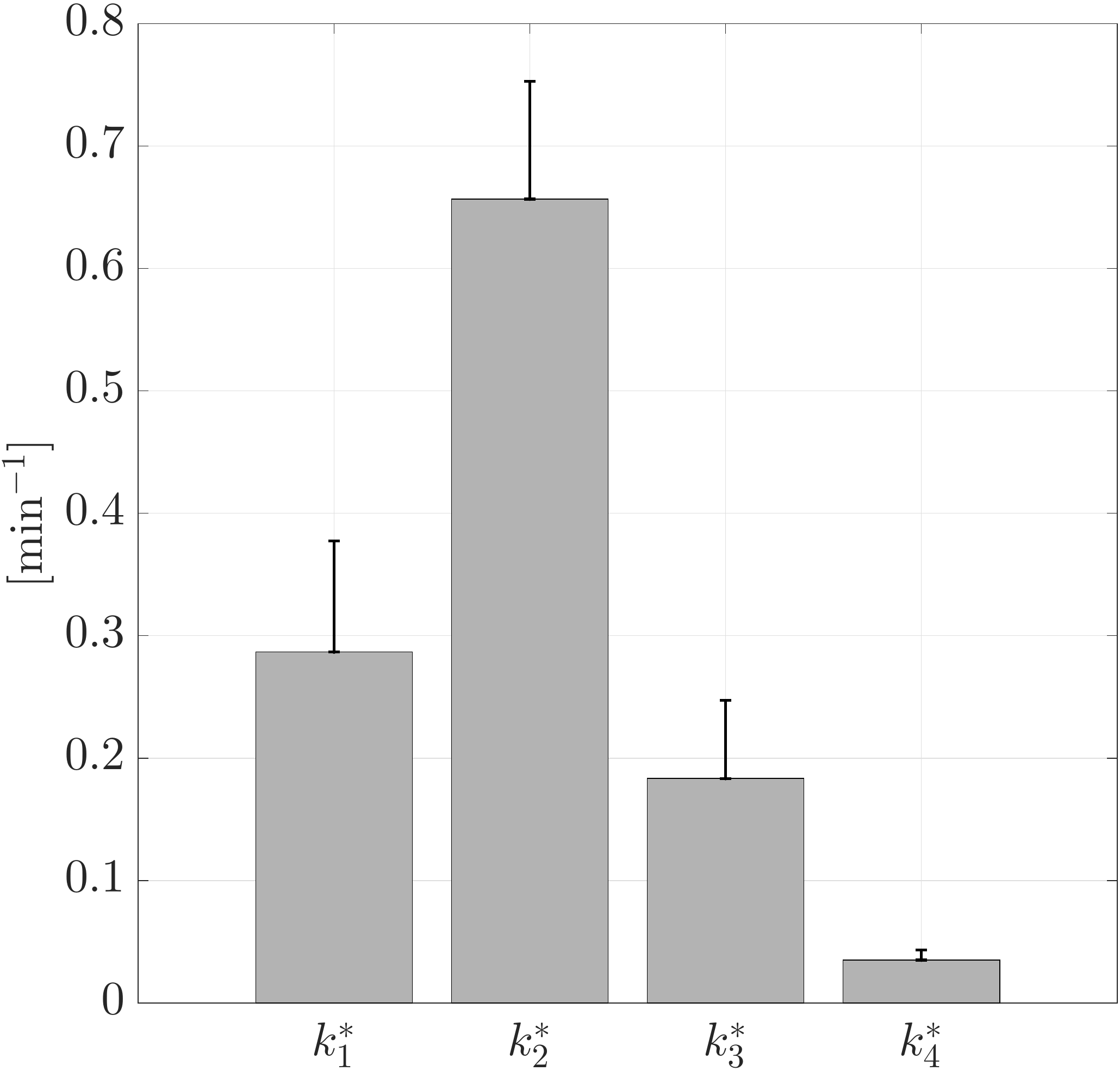}} 
\caption{Bar plot of the reconstructed kinetic parameters: for each model parameter, the mean value and the standard deviation are computed over the mean estimates of the six mouse models for the (a) BCM, and (b) SCM} 
\label{fig:K_bar_tissue}
\end{figure}

A group of six mice, denoted as m$i$, with $i =1,\dots,6$, have been analyzed. The experimental ROI concentration of the CT26 tumor $\mathcal{C}_T$ obtained for one of the mice (specifically, the mouse m1) is shown in \figurename~\ref{fig:CT}; the related canonical arterial IF $C_i$ is shown in \figurename~\ref{fig:Ci}. 
Tissue data have been processed by both the BCM and SCM. Estimates of the parameters obtained for each experiment of the group are reported in \tablename~\ref{tab:K_BCM_tissue} for the BCM, and in \tablename~\ref{tab:K_SCM_tissue} for the SCM. Means and standard deviations have been computed by using 50 runs of the code for the regularized algorithm, with fifty different random initialization values, and with the regularization parameter determined at each iteration through the GCV \citep{GCV} with a confidence interval ranging between $10^4$ and $10^6$. To stop the iterative algorithm we checked the relative error between the experimental concentration and the reconstructed one, computed with the L$^2$ norm, using a threshold of order of $10^{-1}$ as a stopping criterion. For ease of comparison between the reconstructions obtained with the BCM and SCM, \figurename~\ref{fig:K_bar_tissue} shows the bar plot of the kinetic parameters as means and standard deviations computed over the six mouse experiments.
 
To comment on the parameter values reported in \tablename~\ref{tab:K_BCM_tissue}, \tablename~\ref{tab:K_SCM_tissue}, and in \figurename~\ref{fig:K_bar_tissue}, we observe the following.
\begin{itemize}
\item Differences in the estimated values obtained for each mouse agree with the results of simulations of subsection \ref{subsec:PET_simulation}: (1) there is only a slight difference between the reconstructed values of $k_1$, $k_1^*$, and $k_6$, $k_4^*$, respectively; (2) the value of $k_2^*$ is overestimated with respect to $k_2$; (3) $k_3^*$ is underestimated with respect to $k_3$.
\item The parameters $k_6$ and $k_4^*$, related to dephosphorylation, are rather small, but they cannot be neglected since the corresponding fluxes, $k_6 C_r$ and $k_4^* C_p^*$, are comparable with the other contributions; often $k_4^*$ is considered vanishing in Sokoloff-type models, as done by \cite{Roe}, \cite{Rusten}, and \cite{Sokoloff}.
\item Within each table, each parameter is rather stable among the mice of the group, thus showing characteristic kinetic properties of the FDG inside the CT26 tumor tissue, independently of the specific murine experiment.  
\end{itemize}

Estimates of the rate constants available in the literature refer to tissues, and have been obtained by application of compartmental models that are comparable with SCM and are connected to data by equations of the simplified form (\ref{eq:C_T_Sok_vero}), instead of equation (\ref{eq:C_T_tiss_pesato}), which has been applied in the present reduction procedure. The values obtained for $k_1^*$, $k_2^*$, $k_3^*$ and $k_4^*$ are slightly higher than the estimates found, e.g., in \cite{Sokoloff}, referring to cerebral metabolism of albino rats, and in \cite{Reivich} and \cite{Ishibashi}, for cerebral consumption in healthy human brains. 
On the contrary, the present values are comparable with those estimated in \cite{Roe}, for mice with prostate carcinoma xenograft, and in \cite{Rusten}, for soft tissue carcinomas in human patients; in both cases the rate constant corresponding to $k_4^*$ was set equal to zero. 

\begin{figure}[htb]
\centering
\subfigure[BCM. \label{fig:BCM_comp_tissue}]
{\includegraphics[width=5.5cm]{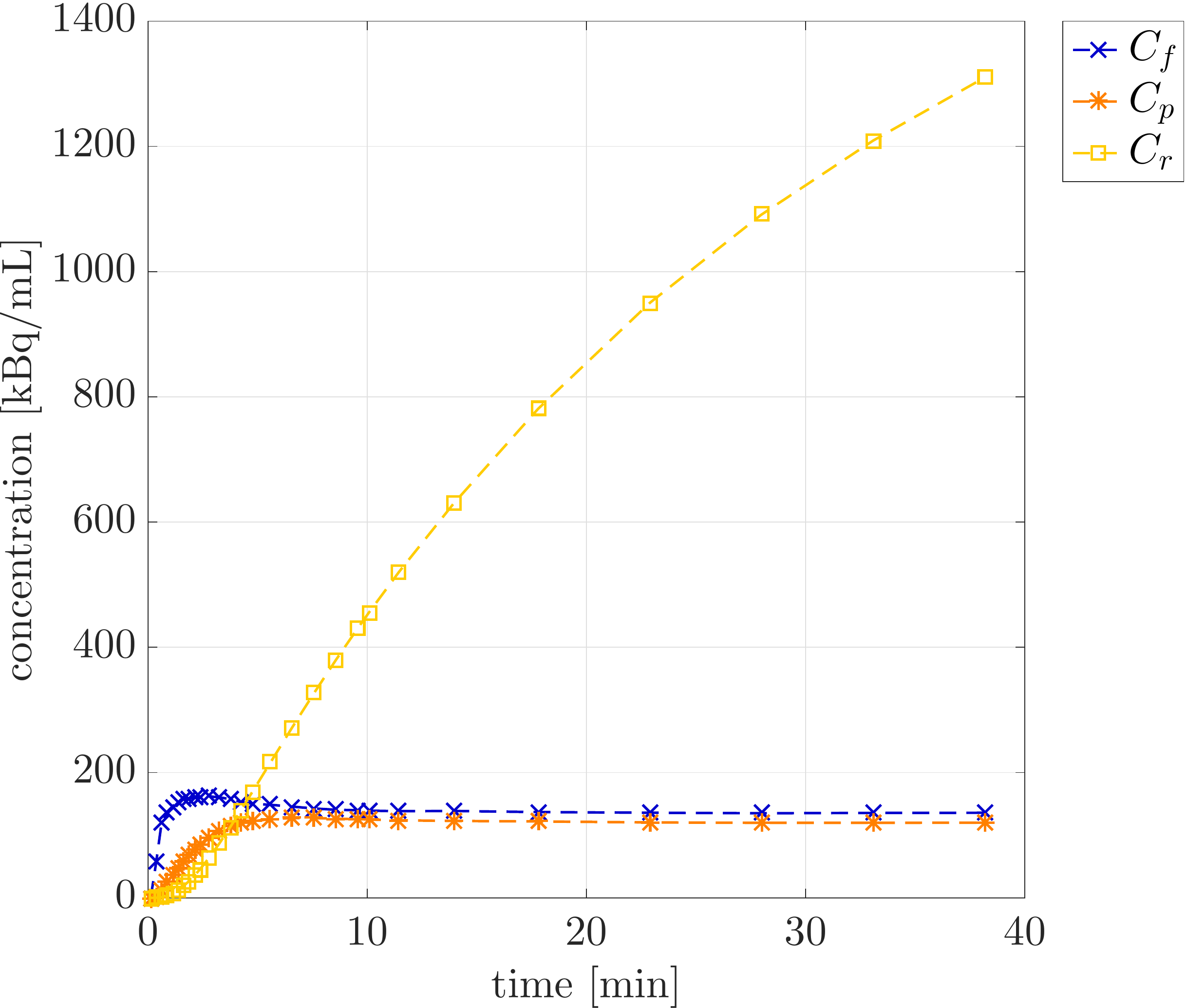}} \ 
\subfigure[SCM. \label{fig:SCM_comp_tissue}]
{\includegraphics[width=5.5cm]{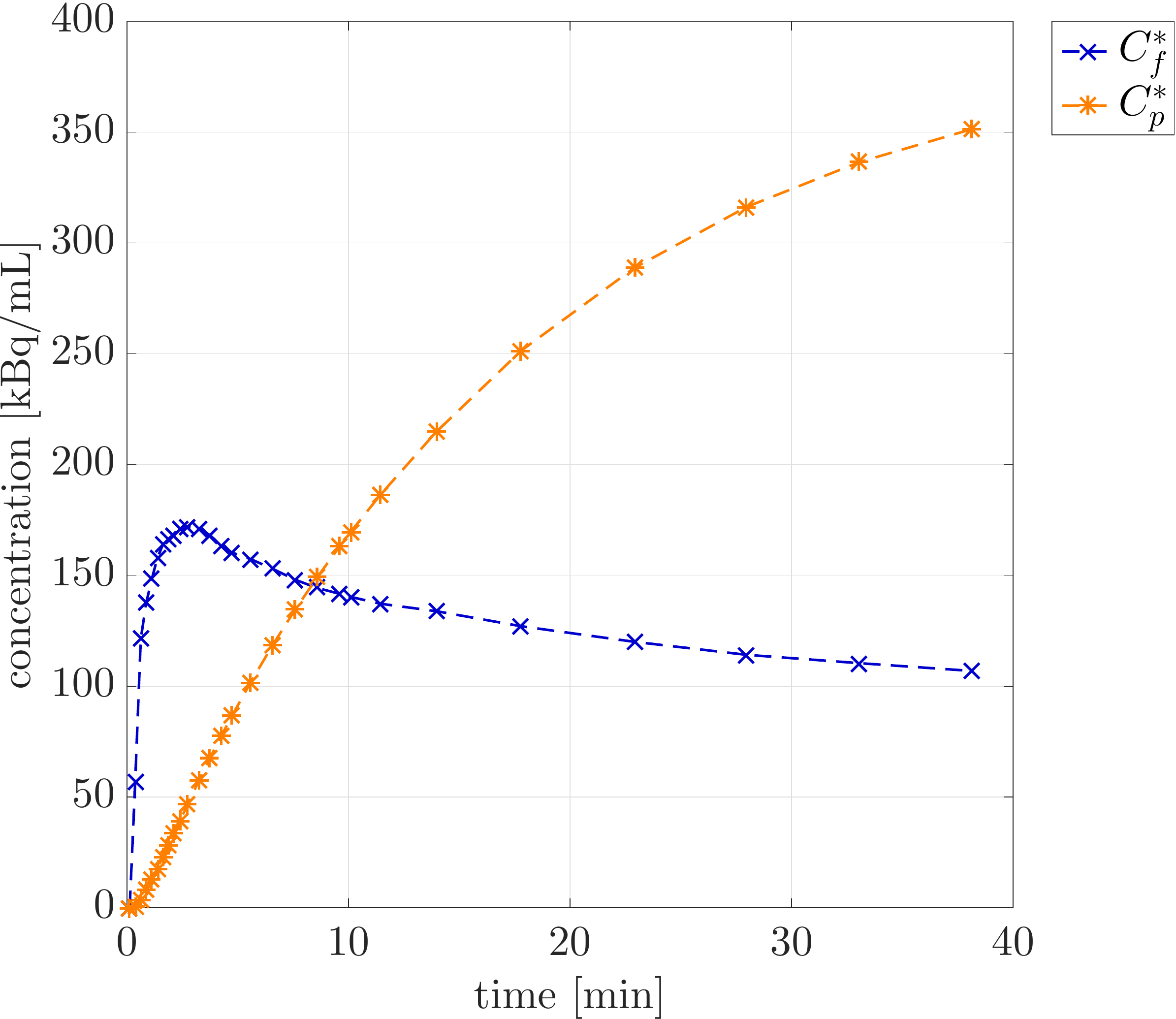}} 
\caption{Model-predicted time curves of the compartment concentrations for the mouse m1: (a) $C_f$, $C_p$ and $C_r$ of the BCM; (b) $C_f^*$, and $C_p^*$ of the SCM} 
\label{fig:comp_tissue}
\end{figure}

Substitution of the values of the rate constants into the systems of ODEs (\ref{eq:ode_C}) and (\ref{eq:dot_C_S}) allows a more complete analysis of tracer kinetics. \figurename~\ref{fig:comp_tissue} shows the reconstructed compartment concentrations for the mouse m1 according to BCM, panel (a), and SCM, panel (b). The curves are representative of the kinetics of the other mice of the group analyzed. 
As to comparison of results obtained for concentrations, it is found that $C_f$ and $C_f^*$ are almost equal for the two models and reach the stationary value in a rather short time. Also, for the BCM, stationarity is achieved by $C_p$ in a few minutes, and its stationary value is smaller than the value of $C_f$. 
According to BCM, tracer accumulates in ER, whose compartment concentration $C_r$ grows with time. Similarly, the analysis performed with the use of SCM shows that accumulation of tracer takes place in the cytosolic phosphorylated pool $C_p^*$. 
Notice that, at each time $t$, $C_r(t)$ is almost four times $C_p^*(t)$. 
This observed difference follows from the fact that the ER compartment in BCM occupies a different volume with respect to the cytosolic phosphorylated compartment in SCM, specifically a smaller volume. Indeed, the activities corresponding to $C_r$ and $C_p^*$ are almost equal, as expected. 
 
\begin{table}
\caption{Reconstructed kinetic parameters (1/min) by the use of the BCM for the CT26 tumor tissue of the FDG--PET experimental group of six mice, as mean and standard deviation over 50 runs of the Gauss-Newton algorithm}
\label{tab:K_BCM_tissue}
\begin{tabular}{llllll}
\hline\noalign{\smallskip}
& $k_1$ & $k_2$ & $k_3$ & $k_5$ & $k_6$ \\
\noalign{\smallskip}\hline\noalign{\smallskip}
m1 & $0.32 \pm 0.03$ & $0.37 \pm 0.15$ & $0.45 \pm 0.19$ & $0.51 \pm 0.28$ & $0.03 \pm 0.02$ \\ 
m2 & $0.47 \pm 0.04$ & $0.67 \pm 0.14$ & $0.54 \pm 0.16$ & $0.59 \pm 0.26$ & $0.03 \pm 0.04$ \\  
m3 & $0.17 \pm 0.03$ & $0.34 \pm 0.17$ & $0.58 \pm 0.21$ & $0.56 \pm 0.25$ & $0.03 \pm 0.02$ \\
m4 & $0.25 \pm 0.03$ & $0.22 \pm 0.12$ & $0.64 \pm 0.21$ & $0.58 \pm 0.23$ & $0.08 \pm 0.02$ \\
m5 & $0.30 \pm 0.04$ & $0.33 \pm 0.19$ & $0.85 \pm 0.31$ & $0.61 \pm 0.27$ & $0.09 \pm 0.05$ \\ 
m6 & $0.31 \pm 0.03$ & $0.36 \pm 0.12$ & $0.61 \pm 0.21$ & $0.53 \pm 0.27$ & $0.09 \pm 0.03$ \\   
\noalign{\smallskip}\hline
\end{tabular}
\end{table} 
    
\begin{table}
\caption{Reconstructed kinetic parameters (1/min) by the use of the SCM for the CT26 tumor tissue of the FDG--PET experimental group of six mice, as mean and standard deviation over 50 runs of the Gauss-Newton algorithm}
\label{tab:K_SCM_tissue}
\begin{tabular}{lllll}
\hline\noalign{\smallskip}
& $k_1^*$ & $k_2^*$ & $k_3^*$ & $k_4^*$ \\
\noalign{\smallskip}\hline\noalign{\smallskip}
m1 & $0.32 \pm 0.02$ & $0.62 \pm 0.09$ & $0.13 \pm 0.09$ & $0.03 \pm 0.04$ \\ 
m2 & $0.43 \pm 0.02$ & $0.84 \pm 0.06$ & $0.11 \pm 0.01$ & $0.03 \pm 0.01$ \\  
m3 & $0.16 \pm 0.03$ & $0.60 \pm 0.19$ & $0.14 \pm 0.06$ & $0.03 \pm 0.02$ \\
m4 & $0.23 \pm 0.02$ & $0.57 \pm 0.11$ & $0.26 \pm 0.14$ & $0.04 \pm 0.03$ \\
m5 & $0.28 \pm 0.02$ & $0.67 \pm 0.12$ & $0.23 \pm 0.05$ & $0.03 \pm 0.01$ \\ 
m6 & $0.30 \pm 0.02$ & $0.64 \pm 0.09$ & $0.23 \pm 0.05$ & $0.05 \pm 0.01$ \\   
\noalign{\smallskip}\hline
\end{tabular}
\end{table}         

\section{Comments and Conclusions} 

In this paper we have examined a new biochemically-driven compartmental model (BCM) aiming at the reconstruction of FDG kinetics in cell cultures and tissues by means of the time dependent input function and total amount of tracer uptake, measured in activity or concentration. 
The BCM originates from an effort of reproducing basic features of tracer kinetics in a single cell; in particular it emphasizes the role of the endoplasmic reticulum, where dephosphorylation of tracer occurs, which had not been considered in previous reconstructions. 
To this aim, an additional compartment for phosphorylated tracer in ER has been introduced, besides the two standard compartments for free and phosphorylated tracer in cytosol. 

The new model has been adapted to the analysis of real data coming from six experiments on cancer cell cultures \emph{in vitro}, and six cancer tissues \emph{in vivo}, also to test its feasibility and effectiveness.
In fact, the framework of cell cultures is naturally related to the cell origin of the model; the tissue framework is related to clinical and physiological applications and allows comparison with results available in literature. 
A simplified version of BCM, referred to as SCM, has been described, which is essentially coincident with a largely used compartmental model, also called Sokoloff model. A second kind of test has been performed, by contrasting the kinetics obtained from the reduction of the same data (on cells and tissues) by application of the two alternative models, BCM and SCM. 
We have found  rather strong similarities in the process of FDG uptake, and significant differences in (1) the reconstructed values of the phosphorylation rate constant ($k_3$ for BCM higher than $k_3^*$ for SCM) and (2) the compartment where accumulation of radioactive tracer occurs (ER for BCM, cytosol for SCM). 

From the modeling viewpoint, the direct connection of BCM to cell biology has been explicitly discussed. It has been found that tracer kinetics is described in terms of five rate constants, instead of the usual four. Then, the cell-based three-compartment BCM has been successfully extended to the analysis of data coming from \emph{in vitro} and \emph{in vivo} measurements. In order to connect the formal model compartments with the available data, additional parameters related to, e.g., cell and tissue physiologies, have been introduced in the model; they have been regarded as given, in order to concentrate on applicability of BCM under natural conditions. 

From a mathematical perspective, it has been shown that the BCMs for cells and tissues are identifiable under rather general conditions. This means that the rate constants are uniquely defined as solutions of the inverse problem, under the assumption that the data are free of noise. 
A detailed comparison has also been made on the kinetic parameters obtained by application of BCM and SCM to the reduction of the same data; it is shown in particular that the two alternative set of values of the rate constants must satisfy an equation that may be regarded as an \emph{a priori} constraint. Slightly different constraint equations hold for cell cultures and tissues.  

From the viewpoint of the data, we observe that deeply different features characterize the cell and tissue systems analyzed in this work. 
(1) The IFs of the cell systems are almost constant, while the IFs of tissue systems show a sharp peak at the initial time. 
(2) The datum of tracer assumption has been reconstructed through highly differing processes based on direct measurement of the radiation emitted (LT) and analysis of reconstructed images (PET image data).
(3) Cell cultures and tissues are inserted in different environments (clean incubation medium vs heterogeneous background, including blood and interstitial tissue), are constituted by different type of cancer cells (4T1 vs CT26), and occupy different total volumes. However, despite all these discrepancies, reconstructed kinetics have shown rather ``stable'' characteristics with respect to the two different biological systems. The similar performance of BCM in such different environments is thus a further strong indication of its reliability. 

From the viewpoint of the new results obtained by application of the BCM, we point out that our conclusions come from analysis of real data. Tracer is shown to accumulate in phosphorylated form in the ER compartment, both for cancer cells and cancer tissues; only a relatively small amount of phosphorylated tracer is found outside the ER. This result follows only from an analysis where, in principle, the two available pools for phosphorylated tracer have been treated on the same level; in this sense, accumulation of tracer in ER is a direct consequence of the inversion procedure and of the properties of data.
We recall that the result has been confirmed in the forthcoming paper (Scussolini et al. manuscript in preparation) by direct measurement on cells seeded \emph{in vitro} and immersed in fluorescent FDG-analogue tracer NBDG.
The value of the phosphorylation rate constant $k_3$, estimated by application of BCM to tissue data (about 0.6 min$^{-1}$), is greater than $k_3^*$ of the SCM (about 0.18 min$^{-1}$), and agrees with results of direct measurements reconstructed from the literature \citep{Gao,Muzi}. Moreover, $k_3^*$ is higher than, or comparable to, estimates of this parameter obtained in literature by application of Sokoloff-type compartmental models \citep{Ishibashi,Reivich,Roe,Rusten,Sokoloff}.
This shows that the phosphorylation rate has been underestimated, and that the proposed BCM gives rise to realistic results. 

As to future tissue applications, the basic scheme of BCM is rather general and may be modified to allow for consideration of specific organs, as done for the SCM in \cite{Garbarino_kidney,Garbarino_liver}, may be associated with reference tissue formulations \citep{Scusso_RTM} or to pixel-wise analysis \citep{Scusso_IP}.
As to cell cultures, an immediate application of BCM is obtained when the composition of the incubation medium is changed in order to examine effects induced on FDG (and perhaps glucose) consumption. 
As to cell biology, the transport of FDG towards the ER is likely to be paralleled by similar migration of glucose in the same direction. Together with the competition between FDG and glucose for cell transmembrane transport and entrapment mechanisms, the role of the ER in the glucose metabolism needs for further investigation. 

Although endowed with new realistic features, the BCM is to be regarded as a simplification with respect to the effective biochemical path followed by FDG inside cells. Nevertheless, the results obtained show that BCM may represent the starting point for the development of a finer and more detailed model able to depict faithfully the FDG and glucose destiny in cancer cells.

 



\end{document}